\documentclass[preprint,secnumarabic,amssymb,superscriptaddress,nobibnotes,aps,prl]{revtex4}

\newcommand{\env}[1]{\texttt{#1}}
\usepackage[version=3]{mhchem}
\usepackage{epstopdf}
\usepackage{braket}
\usepackage{graphicx}
\usepackage{amsfonts,amssymb,amsmath}
\usepackage {mathrsfs}
\usepackage{pxfonts,txfonts}
\usepackage{textcomp}
\graphicspath{{./fig/}}

\begin{document}

\title{Evidence for a Pressure-induced Phase Transition of Few-layer Graphene to 2D Diamond}

\author{Luiz G. Pimenta Martins}
\email{lmartins@mit.edu}
\affiliation {Physics Department, Massachusetts Institute of Technology, Cambridge, MA 02139}

\author{Diego L. Silva}
\affiliation{Departamento de F\'{\i}sica, Universidade Federal de Minas Gerais, Belo Horizonte, MG 30123-970, Brazil}
\author{Jesse S. Smith}
\affiliation{High Pressure Collaborative Access Team, X-ray Science Division,Argonne National Laboratory, Argonne, Illinois 60439, USA}
\author{Ang-Yu Lu}
\affiliation {Department of Electrical Engineering and Computer Science, Massachusetts Institute of Technology, Cambridge, MA 02139}
\author{Cong Su}
\affiliation {Department of Nuclear Science and Engineering, Massachusetts Institute of Technology, Cambridge, MA 02139}
\author{Marek Hempel}
\affiliation {Department of Electrical Engineering and Computer Science, Massachusetts Institute of Technology, Cambridge, MA 02139}
\author{Connor Occhialini}
\affiliation {Physics Department, Massachusetts Institute of Technology, Cambridge, MA 02139}
\author{Xiang Ji}
\affiliation {Department of Electrical Engineering and Computer Science, Massachusetts Institute of Technology, Cambridge, MA 02139}
\author{Ricardo Pablo}
\affiliation {Department of Nuclear Science and Engineering, Massachusetts Institute of Technology, Cambridge, MA 02139}
\author{Rafael S. Alencar}
\affiliation {Departamento de F\'{\i}sica, Universidade Federal de Minas Gerais, Belo Horizonte, MG 30123-970, Brazil}
\author{Alan C. R. Souza}
\affiliation{Departamento de F\'{\i}sica, Universidade Federal de Minas Gerais, Belo Horizonte, MG 30123-970, Brazil}
\author{Alan B. de Oliveira}
\affiliation{Departamento de F\'{\i}sica, Universidade Federal de Ouro Preto, Ouro Preto, MG 35400-000, Brazil, Brazil}
\author{Ronaldo J.C. Batista}
\affiliation{Departamento de F\'{\i}sica, Universidade Federal de Ouro Preto, Ouro Preto, MG 35400-000, Brazil, Brazil}
\author{Tom\'as Palacios}
\affiliation{Department of Electrical Engineering and Computer Science, Massachusetts Institute of Technology, Cambridge, MA 02139}
\author{Matheus J.S. Matos}
\affiliation{Departamento de F\'{\i}sica, Universidade Federal de Ouro Preto, Ouro Preto, MG 35400-000, Brazil, Brazil}
\author{M\'ario S.C. Mazzoni}
\affiliation{Departamento de F\'{\i}sica, Universidade Federal de Minas Gerais, Belo Horizonte, MG 30123-970, Brazil}
\author{Riccardo Comin}
\affiliation{Physics Department, Massachusetts Institute of Technology, Cambridge, MA 02139}
\author{Jing Kong}
\email{jingkong@mit.edu}
\affiliation{Department of Electrical Engineering and Computer
Science, Massachusetts Institute of Technology, Cambridge, MA 02139}
\author{Luiz G. Can\c{c}ado}
\email{cancado@fisica.ufmg.br}
\affiliation{Departamento de F\'{\i}sica, Universidade Federal de Minas Gerais, Belo Horizonte, MG 30123-970, Brazil}

\begin{abstract}
We unveil the diamondization mechanism of few-layer graphene compressed in the presence of water, providing robust evidence for the pressure-induced formation of 2D diamond. High-pressure Raman spectroscopy provides evidence of a phase transition occurring in the range of 4--7\,GPa for 5--layer graphene and graphite. The pressure-induced phase is partially transparent and indents the silicon substrate. Our combined theoretical and experimental results indicate a gradual top-bottom diamondization mechanism, consistent with the formation of diamondene, a 2D ferromagnetic semiconductor. High-pressure x-ray diffraction on graphene indicates the formation of hexagonal diamond, consistent with the bulk limit of eclipsed-conformed diamondene.
\end{abstract}
\maketitle

The search for a stable 2D diamond has gathered recent interest due to the possibility of combining diamond's distinguished properties, such as superior hardness~\cite{zhao2016recent} and heat conduction~\cite{balandin2011thermal}, with exotic new properties that may arise from the reduced dimensionality. Its existence was first proposed over a decade ago~\cite{chernozatonskii2009diamond},
and different structures have been theoretically proposed ever since~\cite{chernozatonskii2010influence,barboza2011room,antipina2015converting,gao2018ultrahard}. In most structures, stability is achieved by surface functionalization at the top and bottom surfaces~\cite{chernozatonskii2009diamond,antipina2015converting,gao2018ultrahard}, sometimes called diamane~\cite{chernozatonskii2009diamond} for bilayer, and diamano\"{i}ds for thicker layers~\cite{piazza2019low}. Diamondene\cite{martins2017raman}, another proposed 2D--diamond structure, greatly differs from those, being covalently bonded to chemical groups only at the top surface while the bottom exhibits a periodic array of dangling bonds. These unpaired electrons generate magnetism in diamondene, and their periodic distribution gives rise to two spin-polarized bands, making it an ideal platform material for spintronics~\cite{barboza2011room,martins2017raman}.\\

In this letter, we investigate the formation of diamondene via high-pressure experiments using diamond anvil cells (DACs) and we provide structural information through high-pressure x-ray diffraction (XRD). We use water as the pressure transmitting medium (PTM) due to its importance in facilitating this phase transition~\cite{barboza2011room,martins2017raman}. High-pressure Raman spectroscopy provides evidence of a phase transition occurring in the range of 4--7\,GPa for 5-layer graphene and graphite. The pressure-induced phase is partially transparent and indents the silicon substrate, indicative of extreme hardness. We combine experimental data with Molecular dynamics (MD) simulations and Density functional theory (DFT) calculations to propose a full diamondization mechanism in compressed few-layer graphene, providing robust evidence for the pressure-induced formation of 2D diamond. We show that the use of water PTM allows for a gradual top-bottom diamondization process, consistent with the formation of diamondene.\\

\begin{figure}[!tb]
\begin{center}
\includegraphics [scale=0.18]{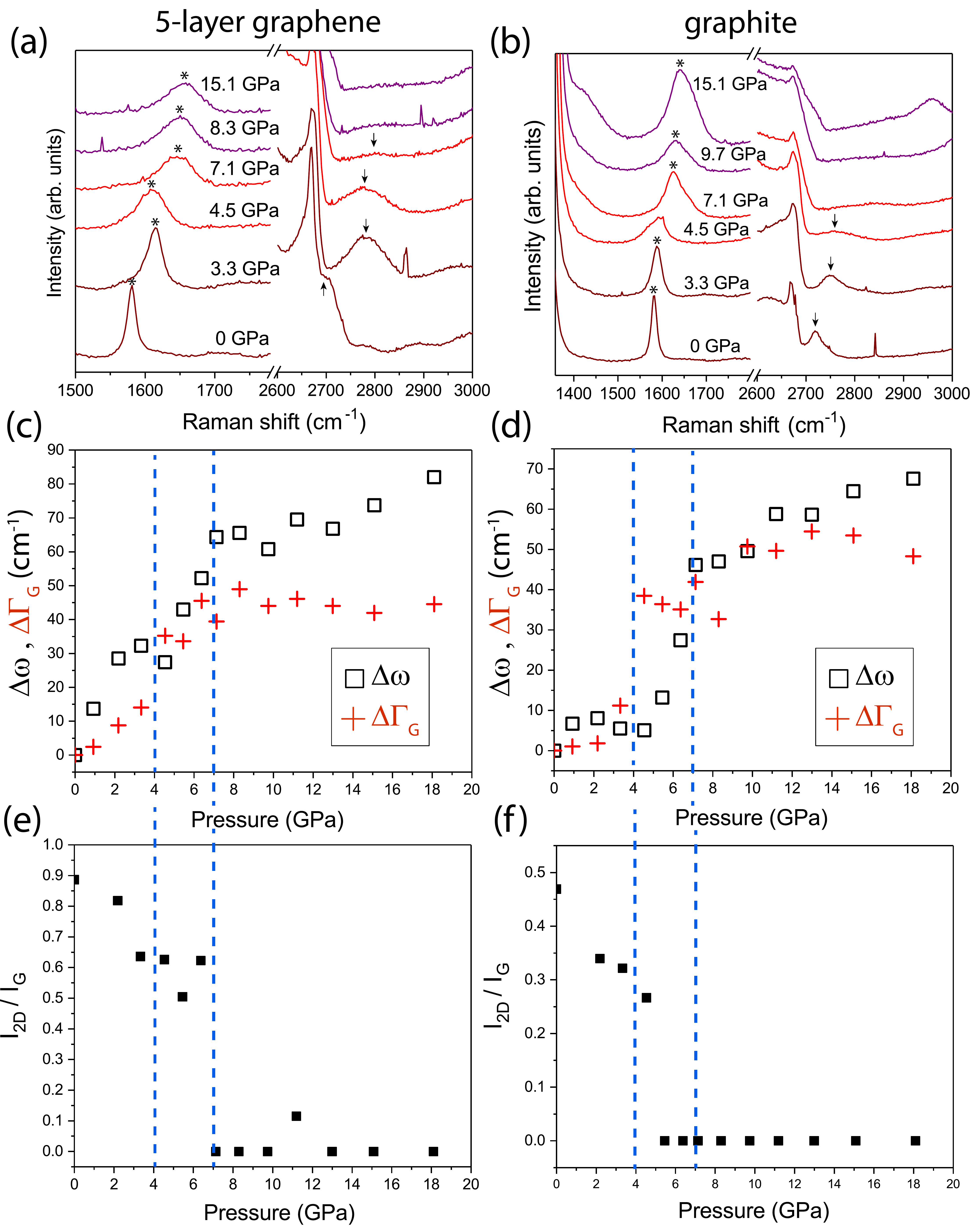}
\caption{\label{fig2} Phase transition evidence obtained from Raman spectroscopy. (a,b) Selected Raman spectra at different pressures for (a) five-layer graphene and (b) graphite. G and 2D bands are indicated with $\ast$ and $\downarrow$, respectively. (c,d) Plots of the G band frequency and full-width at half maximum subtracted from their values at initial pressure, $\Delta \omega_{\rm G}$ and $\Delta \Gamma_{\rm G}$, respectively, as a function of pressure for (c) five-layer graphene and (d) graphite. The blue-dashed lines indicate the critical pressures of the phase transition. (e,f) Plots of the intensity ratio between the 2D and G band ($I_{\rm 2D}/I_{\rm G}$) for (e) five-layer graphene, and (h) graphite.}
\end{center}
\end{figure}

High-pressure Raman experiments were performed on mechanically exfoliated graphene samples transferred onto 25-$\mu$m-thick Si substrates (see details in Supplementary Information). The sample is composed of a graphite piece sitting next to a five-layer graphene, as shown in Fig.~S1(d). The sample was compressed in the DAC using water as the PTM, and all Raman spectra were acquired using a 532\,nm excitation laser. Figures~\ref{fig2}(a,b) show selected Raman spectra of the compressed five-layer graphene and graphite, respectively, measured at different pressures featuring the first-order Raman-allowed G band ($\approx$\,1580\,cm$^{-1}$ in ambient pressure), and the two-phonon Raman 2D band ($\approx$\,2700\,cm$^{-1}$)~\cite{jorio2011raman}.\\

Figures~\ref{fig2}(c,d) show the plot of the pressure-evolution of the G band frequency subtracted from its frequency measured in ambient conditions ($\Delta\omega_{\rm G}$). The data were extracted from the Raman spectra of five-layer graphene and graphite, respectively. Figures~\ref{fig2}(c,d) show the evolution of the full-width at half maximum of the G band, $\Delta\Gamma_{\rm G}$. At low pressures, we observe the expected strain-induced phonon hardening~\cite{huang2009phonon}, by shifts in the G and 2D bands~\cite{proctor2009high} to higher frequencies without significant changes in $\Gamma_{\rm G}$~\cite{filintoglou2013raman}. The trend changes at $\approx$\,4 and $\approx$\,7\,GPa for both flakes. These critical pressures are evidenced by the blue dashed lines in Figs.~\ref{fig2}(c,d). The most remarkable change is an abrupt increase in $\Gamma_{\rm G}$, which is considered a signature of structural phase transitions in compressed graphite~\cite{hanfland1989graphite,amsler2012crystal,wang2012crystal}. For both five--layer graphene [Figs.~\ref{fig2}(a,c)] and graphite [Figs.~\ref{fig2}(b,d)], $\Gamma_{\rm G}$ suddenly broadens around 4\,GPa, indicating the onset of a phase transition, and remains relatively constant above 7\,GPa, indicating a completed transition.\\ 

\begin{figure}[!tb]
\begin{center}
\includegraphics [scale=0.35]{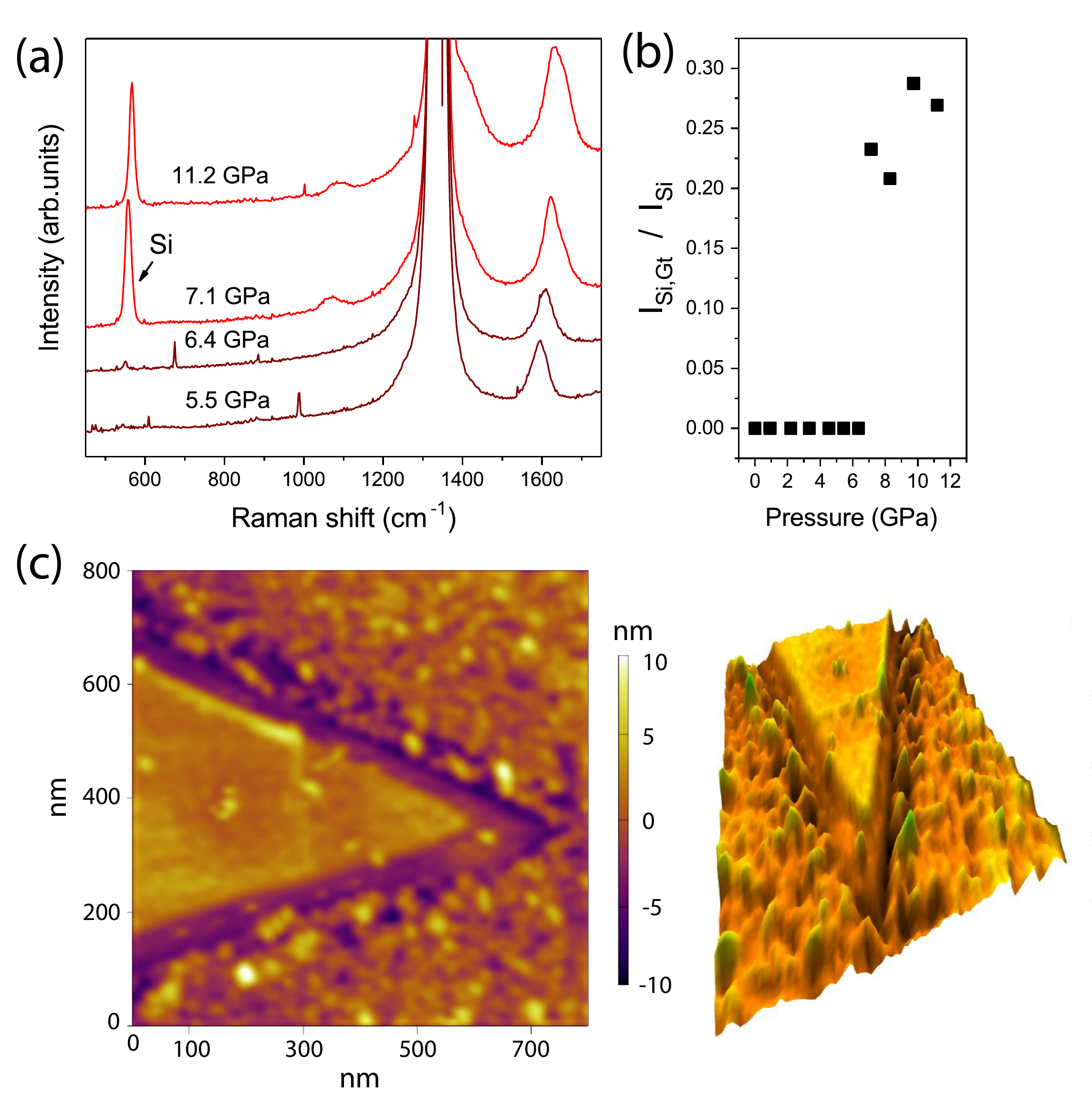}
\caption{\label{fig3} Evidence of pressure-induced transparency and formation of a hard phase. (a) Raman spectra of graphite at different pressures. From 7.1\,GPa, the Raman peak from the silicon substrate appears at approximately 550\,cm$^{-1}$. (b) Plot of the ratio between intensities of the silicon peak from the substrate area under the graphite piece ($I_{\rm Si,Gt}$) and from an uncovered substrate area nearby ($I_{\rm Si}$). (c) AFM topographical images of the graphite flake post compression, showing deep grooves formed along the edges of the flake. The image on the right is a 3D-perspective of the image on the left.}
\end{center}
\end{figure}

The occurrence of a phase transition is further bolstered by the concomitant abrupt intensity suppression of the 2D band, as observed in the Raman spectra shown in Figs.~\ref{fig2}(a,b) and systematically depicted in the plots of the ratio between the 2D and G bands intensities, $I_{\rm 2D}/I_{\rm G}$, as a function of pressure, shown in Figs.~\ref{fig2}(e,f) for the five-layer graphene and graphite, respectively. Since the 2D band originates from double-resonance mechanisms~\cite{jorio2011raman}, its intensity is highly sensitive to changes in the electronic structure due to structural changes~\cite{ferreira2010evolution}.\\ 

After 7.1\,GPa, the Raman peak at $\approx$\,1520\,cm$^{-1}$ from the SiO$_2$/Si substrate area under the graphite flake becomes detectable, as shown in Fig.~\ref{fig3}(a)-- an indication of transparency. Figure~\ref{fig3}(b) quantifies this transparency event by showing the plot of the ratio between the intensity of the Si Raman peak measured from an area under the graphite flake, $I_{\rm Si, Gt}$, and its intensity measured directly on the Si substrate,  $I_{\rm Si}$, as a function of pressure. As Fig.~\ref{fig3}(b) shows, the ratio becomes $\approx\,0.25$ for pressures above 7.1\,GPa, indicating that the flake became partially transparent. Indeed, the optical images of the sample inside of the DAC at different pressures suggest a gradual top-bottom increase in transparency of the flakes in the 4--7\,GPa range (see discussion in Supplementary Information). \\

\begin{figure}[!tb]
\begin{center}
\includegraphics [scale=0.35]{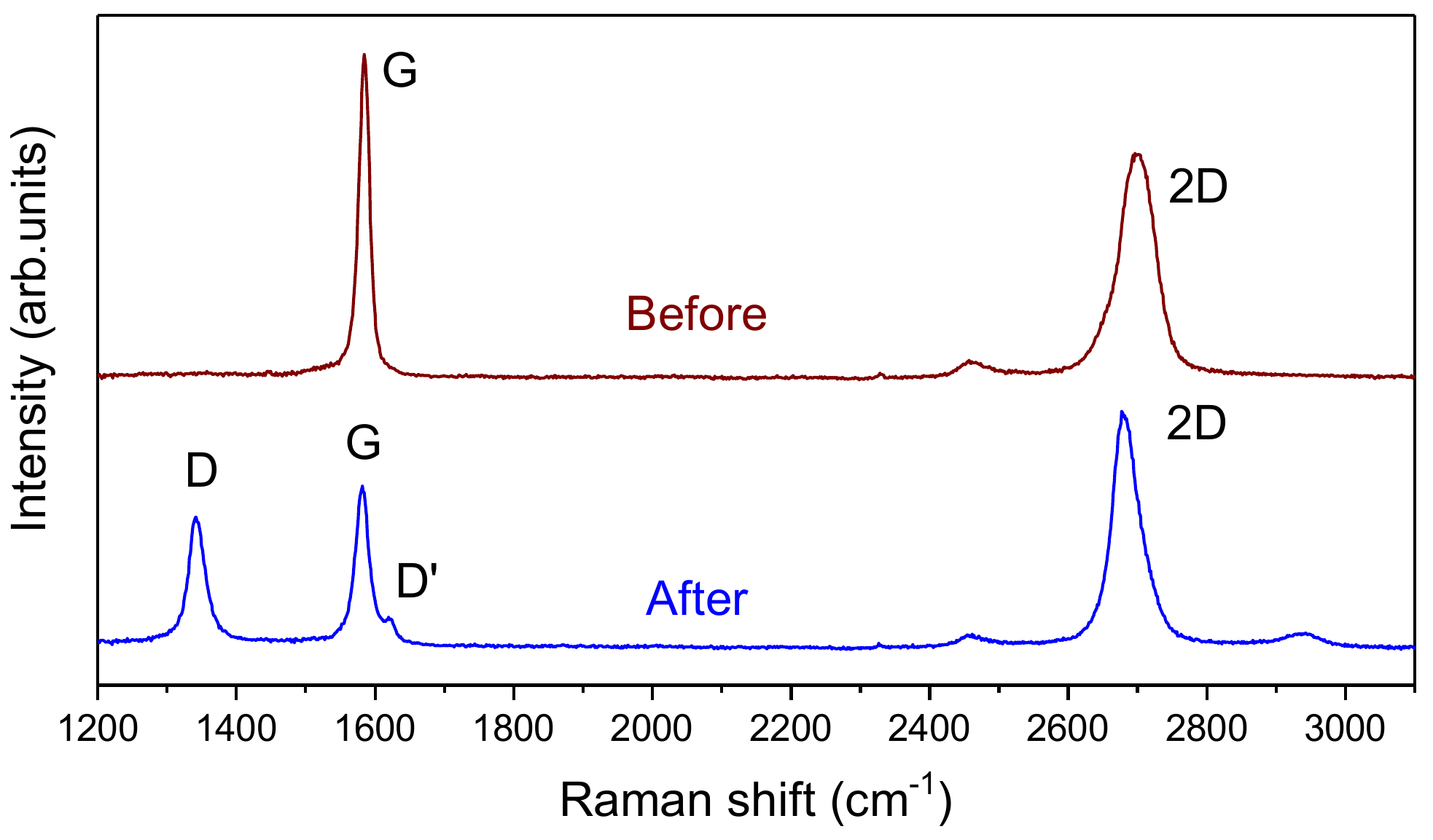}
\caption{\label{fig6} Raman spectra acquired at ambient pressure before and after compression of the four--layer graphene sample.}
\end{center}
\end{figure}

Upon phase transition, the optical images show cracks on the substrate along directions defined by the edges of the flake (Figure~S3), indicating the formation of a hard phase. To confirm this feature we compressed a sample containing a four--layer graphene and a graphite flake on SiO$_2$/Si in a DAC, up to 8\,GPa using water PTM. The atomic force microscopy (AFM) topographical images of the recovered sample clearly show indentation marks on the SiO$_2$, following the shape of both graphite [Fig.~\ref{fig3}(c)] and four--layer sample [Fig.~S4] . Furthermore, the Raman spectra of all recovered samples are remarkably similar to those of hydrogenated graphene~\cite{elias2009control}, exhibiting strong defect-induced D and D$^{'}$ Raman bands~\cite{jorio2011raman}, consistent with a partial functionalization of graphene. As an example, Fig.~\ref{fig6} shows the Raman spectra before and after decompression, at ambient condition, for the four--layer graphene.\\ 

\begin{figure}[!tb]
\begin{center}
\includegraphics [scale=0.35]{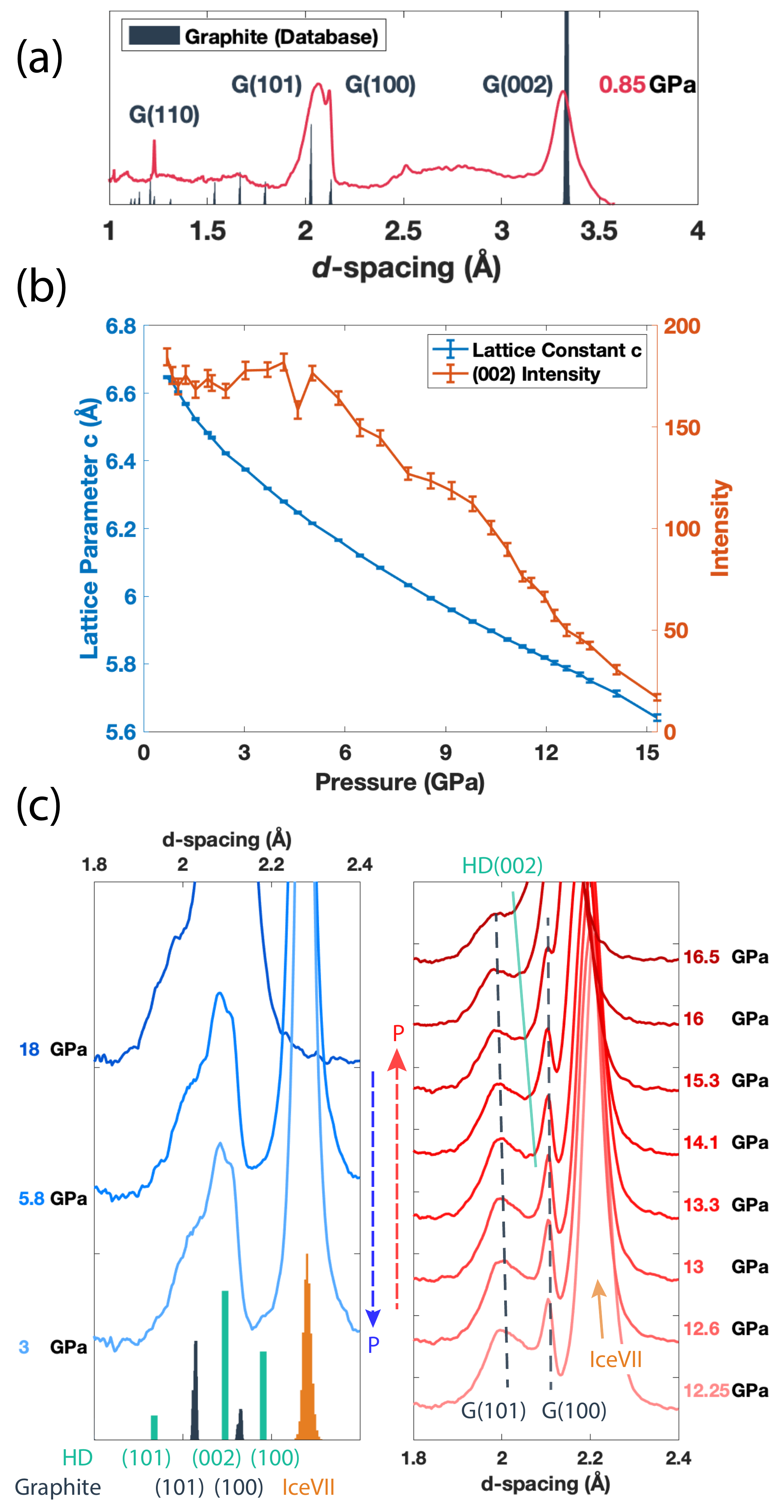}
\caption{\label{fig4} (a) XRD intensity as a function of $d$--spacing of the graphene powder at the initial pressure. The peaks from graphitic/graphene (G) systems at 0\,GPa are labeled. (b) Intensity and $c$ parameter obtained from the G002 peak as a function of increasing pressure. Error bars represent the 95\% confidence interval obtained from the fitting. (c) Zoomed XRD diffractograms upon decompression, from maximum pressure of 18\,GPa down to residual pressure of $\approx$\,3\,GPa as a function of $d$--spacing. The peaks from hexagonal diamond (HD), graphite (G), and Ice VII are labeled at the bottom. The hexagonal diamond peaks were simulated from the bulk limit of BB stacked diamondene at 0\,GPa. These peaks match those reported for hexagonal diamond in Ref.~\cite{yagi1992high}. Upon decompression, the intense Ice VII peak ($d$--spacing $\approx$\,2.1\,\AA\,at 18\,GPa) retracts and a resolvable new peak can be seen at $\approx$\,2.08\,\AA, consistent with HD002, in between G100 and G101. (d) XRD  diffractograms at increasing pressures showing the HD002 peak appearing at $\approx$\,2.06\,\AA\, at approximately 14\,GPa in between G100 and G101.}
\end{center}
\end{figure}

To obtain information on the crystal structure of the new phase, we performed high-pressure XRD on a graphene powder using water PTM. A graphene powder was used instead of a single few-layer graphene crystal, since the latter presents enormous challenges in terms of low x-ray cross section, given its low atomic number and two-dimensional nature. The sample contains $86\%$ by mass of flakes with thickness below 20 layers (see details in Supplementary Information). The sample was compressed to a maximum pressure of 18\,GPa, and decompressed to a final residual pressure of $\approx$\,3\,GPa. Figure~\ref{fig4}(a) shows the diffraction intensity as a function of $d$--spacing at the initial pressure. The peaks from graphitic/graphene (G) systems at 0\,GPa are labeled in Fig.~\ref{fig4}(a).\\

Figure~\ref{fig4}(b) shows the evolution, with increasing pressure, of the intensity and $d$-spacing of the G002 peak, from which the out-of-plane lattice parameter $c$ can be obtained. The G002 peak shifts with increasing pressure due to graphene's large compressibility along the $c$--axis~\cite{yagi1992high}, and its intensity remains practically constant until $\approx$\,5\,GPa, after which it starts to decrease, becoming undetectable above 15\,GPa. Such a decrease can be assigned to the onset of the phase transition, in agreement with the onset pressures observed in the Raman experiments.  Upon decompression [Fig.~\ref{fig4}(c)], a new peak is resolvable between G101 and G100. The rise of a peak with similar features has been reported in room-temperature compressed graphite upon decompression and assigned as 002 of hexagonal diamond (HD)~\cite{yagi1992high}. As shown in Fig.~\ref{fig4}(b), a good agreement exists for the two strongest peaks HD002 and HD100. Other HD peaks could not be identified, likely due to a combination of low intensity, unfavorable spatial orientation of flakes that meet the Bragg condition and overlap with ICE VII or graphene peaks, as can be seen in Fig.~S11. In fact, it is possible to see the rise of the HD002 peak at approximately 14\,GPa [Fig.\ref{fig4}(c)].\\

To interpret the experimental results, we carried out MD simulations and DFT calculations. Both theoretical schemes assume a starting system of an AB--stacked five--layer graphene whose top surface is covered with either --OH or --H groups. These are assumed to originate from water molecules in reactions under high pressures at the surface of graphene~\cite{barboza2011room} (see details in Supplementary Information), and play a fundamental role in both rehybridization and stabilization processes~\cite{paul2019mechanochemistry}. \\

\begin{figure}[!tb]
\begin{center}
\includegraphics [scale=0.145]{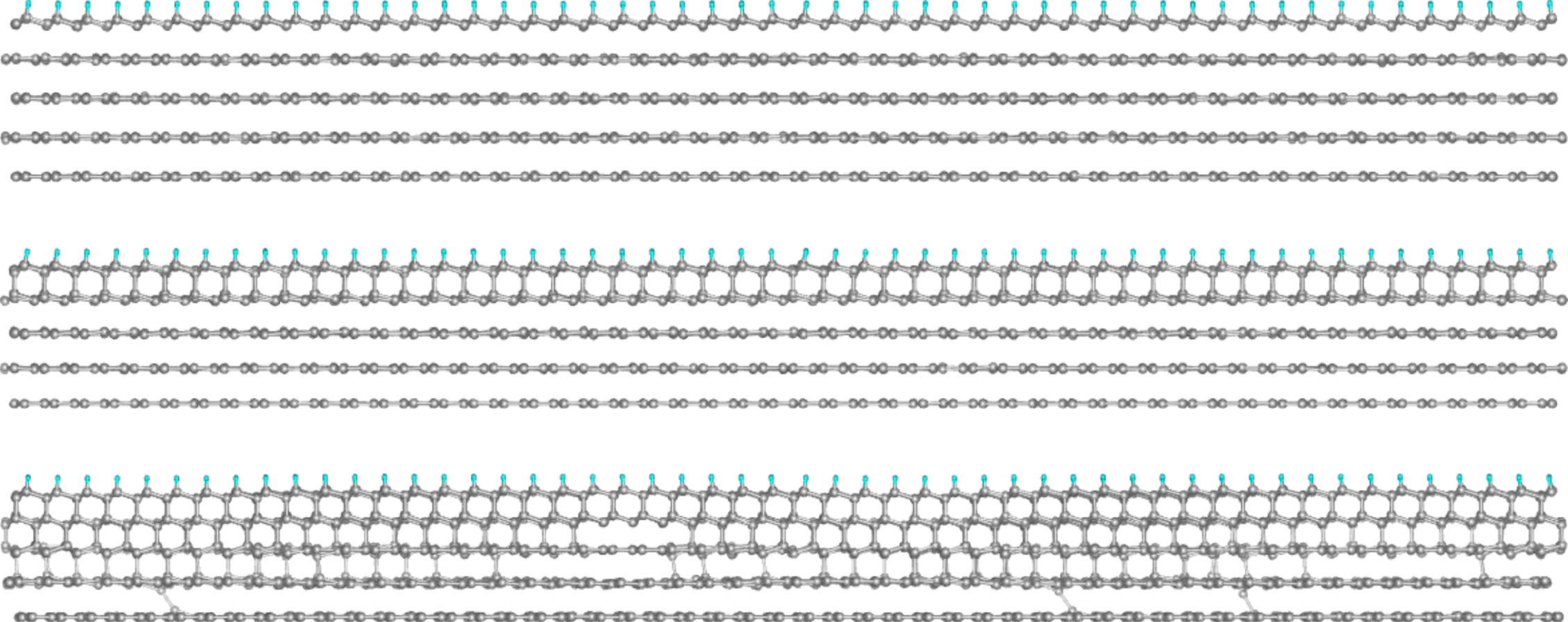}
\caption{\label{fig7} Snapshots of the MD simulation of the H-functionalized five-layer graphene system  with increasing pressure.}
\end{center}
\end{figure}

Figure~\ref{fig7} presents three snapshots with increasing pressure obtained from MD simulations. The pressure was applied to the system by a moving piston modeled as a  Lennard-Jones wall. Upon compression, the phase transition starts with the diamondization of the first two layers, giving rise to diamondene. For the formation of covalent bonds between diamondene and the graphene layer underneath to occur, the pressure needs to be further increased since the distance between the layers is too large to allow for the rehybridization. In this process, called vertical propagation, the third graphene layer displaces horizontally resulting in a transformation from an ABA to an ABB trilayer stacking configuration. \\

\begin{figure}[!tb]
\begin{center}
\includegraphics [scale=0.19]{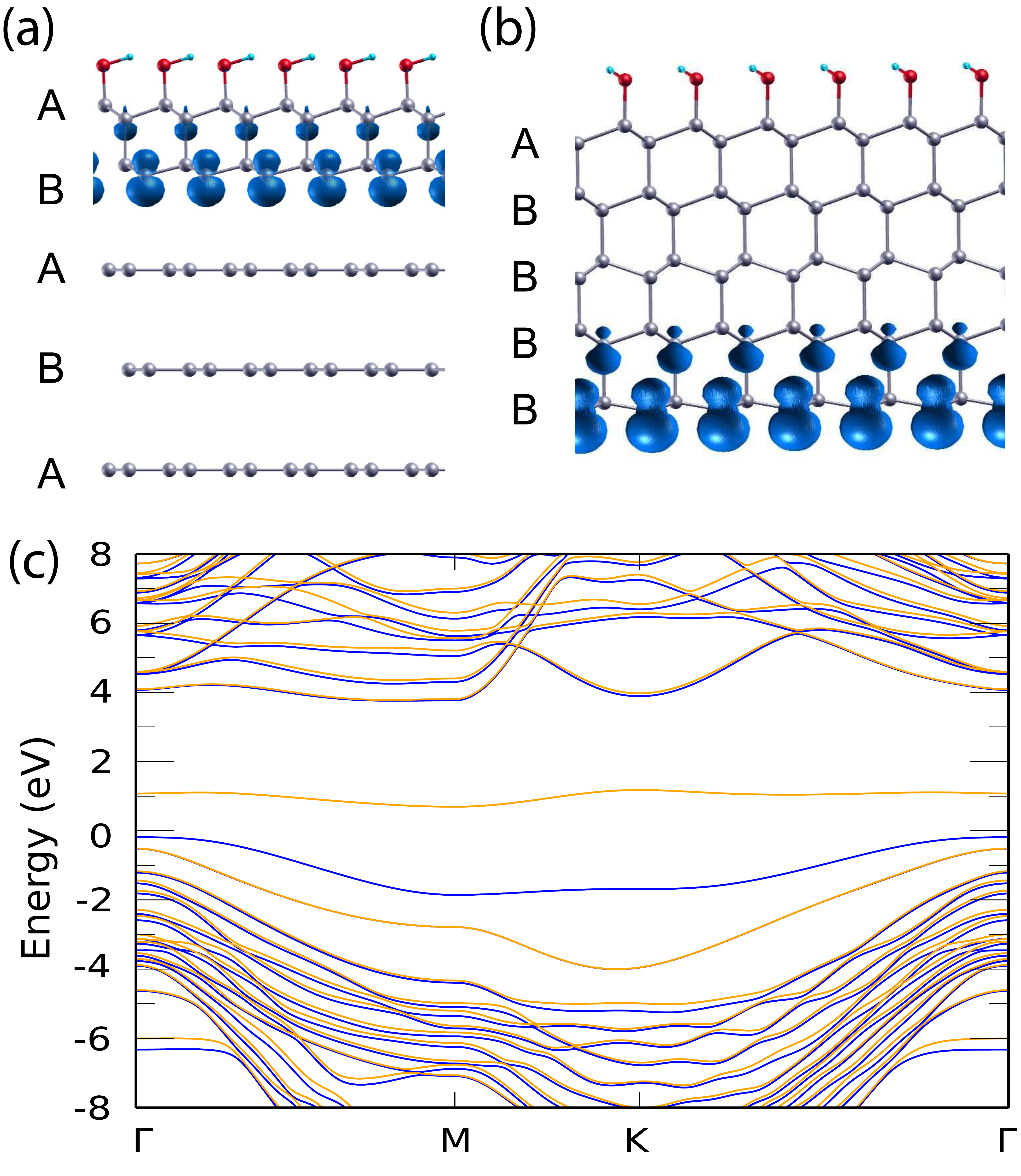}
\caption{\label{fig5} Evolution of the sp$^{2}$--sp$^{3}$ restructuring process. (a,b) DFT results with increasing applied force, also showing the Local Density of States (blue bubbles) of the dangling bonds above Fermi level (see details in Supplementary Information). In the last configuration (b), the system is completely rehybridized, and the staking order is ABBBB. (c) Band structure for the all--sp$^{3}$ configuration. Blue and yellow lines represent the two spin components. }
\end{center}
\end{figure}

The DFT calculations confirm this gradual-top bottom diamodization process. Figures~\ref{fig5}(a,b) show the optimized geometries for two configurations with increasing applied force, obtained from DFT calculations. To mimic the application of pressure, we employed a hard-wall constraint in the lowest carbon atoms, which are not allowed to move in the negative z--direction, while a predefined force was applied in the oxygen atoms of the --OH groups which functionalize the upper graphene surface. Under compression, the diamondene structure is initially formed with the underlying three-layer graphene protecting the array of dangling bonds [Fig.~\ref{fig5}(a)]. Interestingly, a similar structure has been recently reported from a hot-filament process of few-layer graphene~\cite{piazza2019towards}. Next, by increasing the applied forces, a completely rehybridized five--layer geometry characterized by an ABBBB stacking is obtained, as shown in Fig.~\ref{fig5}(b). In the BB part of that stacking, the interlayer bonds are in an eclipsed conformation. The electronic band structure corresponding to the ABBBB stacking, shown in Fig.~\ref{fig5}(c), preserves the main features of the diamondene -- localized states at the dangling bonds leading to a pair of spin-polarized bands. Different rehybridized geometries are possible (see discussion in Supplementary Information).\\

The vertical propagation mechanism explains several features of the experiment, for instance, the XRD data. Since there is no indication of significant water intercalation in the graphene powder- the initial d-spacing is 3.323\,\AA\,[see Fig.\ref{fig4}(b)], while the basal spacing of graphite is 3.3553\,\AA\,\cite{clark2013few}- it is reasonable to propose a top-bottom diamondization process, as shown in Figs.~\ref{fig5}(a,b), starting at the surface (in contact with water), and propagating towards the powder's core. As the crystal grows, in the bulk limit, one should expect a BB stacked configuration [Fig.~\ref{fig5}(b)], which is precisely the crystal structure of hexagonal diamond. Also in the bulk limit, one should expect a detectable XRD signal, with intensity strong enough to overcome the broad graphene peaks. Moreover, the fact that the HD002 peak rises as the G002 graphene peak vanishes is consistent with our model, in which diamondene and graphene share the same $c$-axis.\\

The model also explains the asymmetric shape of the Raman G band for both 5--layer graphene and graphite after the onset of the phase transition. Starting from 4\,GPa, the G band exhibits an asymmetric shape, which is more pronounced for graphite [Figs.~\ref{fig2}(a,b)]. This could be caused by the coexisting pristine and hybridized phases during the vertical propagation [Fig.~\ref{fig5}(a)], with distinct G band frequencies of narrowed and broadened bandwidth, respectively. This feature was confirmed, and other aspects of the model were further investigated in a high-pressure Raman experiment comparing graphite and a graphene powder (see discussion in Supplementary Information). Another important aspect to be noted is the shape of the 2D band of the recovered samples post-compression. As observed from the Raman spectra obtained from the recovered samples, shown in Fig.\ref{fig6}, its symmetrical shape indicates that an overall loss of stacking order occurred in those flakes~\cite{jorio2011raman}, consistent with lateral sliding between layers during compression.\\

Most previous experimental indications of the existence of a 2D diamond have been obtained by tip-compression~\cite{barboza2011room,gao2018ultrahard,cellini2019layer}, hydrostatic compression~\cite{martins2017raman,ke2019large} or hydrogenation of few-layer graphene~\cite{piazza2019low,srivastava2019synthesis}. Even though these works promoted significant advances in the field, they did not provide structural information about the new phase. Recent preprints investigate the formation of diamane~\cite{bakharev2019chemically} and diamano\"{i}ds~\cite{piazza2019towards} via chemical-functionalization routes, reporting structural information through transmission electron microscopy and low-energy electron diffraction, respectively.\\ 

In summary, Raman spectroscopy data identified the critical pressures for five-layer graphene and graphite. The pressure-induced phase is partially transparent and indents the silicon substrate. Our combined theoretical and experimental results indicate a gradual top-bottom diamondization process, which starts with the formation of bi-layer diamondene and propagates along $c$ axis to the bottom. High-pressure XRD data indicated the formation of hexagonal diamond, consistent with the bulk limit of eclipsed-conformed diamondene. Evidence of functionalization from the recovered samples suggests that a stable 2D diamond could be obtained through further tuning of the synthesis parameters, e.g., by increasing temperature.\\ 

\section{Supplementary Information}

\renewcommand{\thefigure}{S\arabic{figure}}
\setcounter{figure}{0}


\subsection{Sample preparation}\label{SP}

To miniaturize the samples, a method based on etching horseshoe-shaped trenches on a thin silicon wafer was developed. This method can be used to systematically load 2D materials and related heterostructures into DACs, a common bottleneck in high-pressure experiments involving these types of systems.\\

\begin{figure}[!tb]
\begin{center}
\includegraphics [scale=0.47]{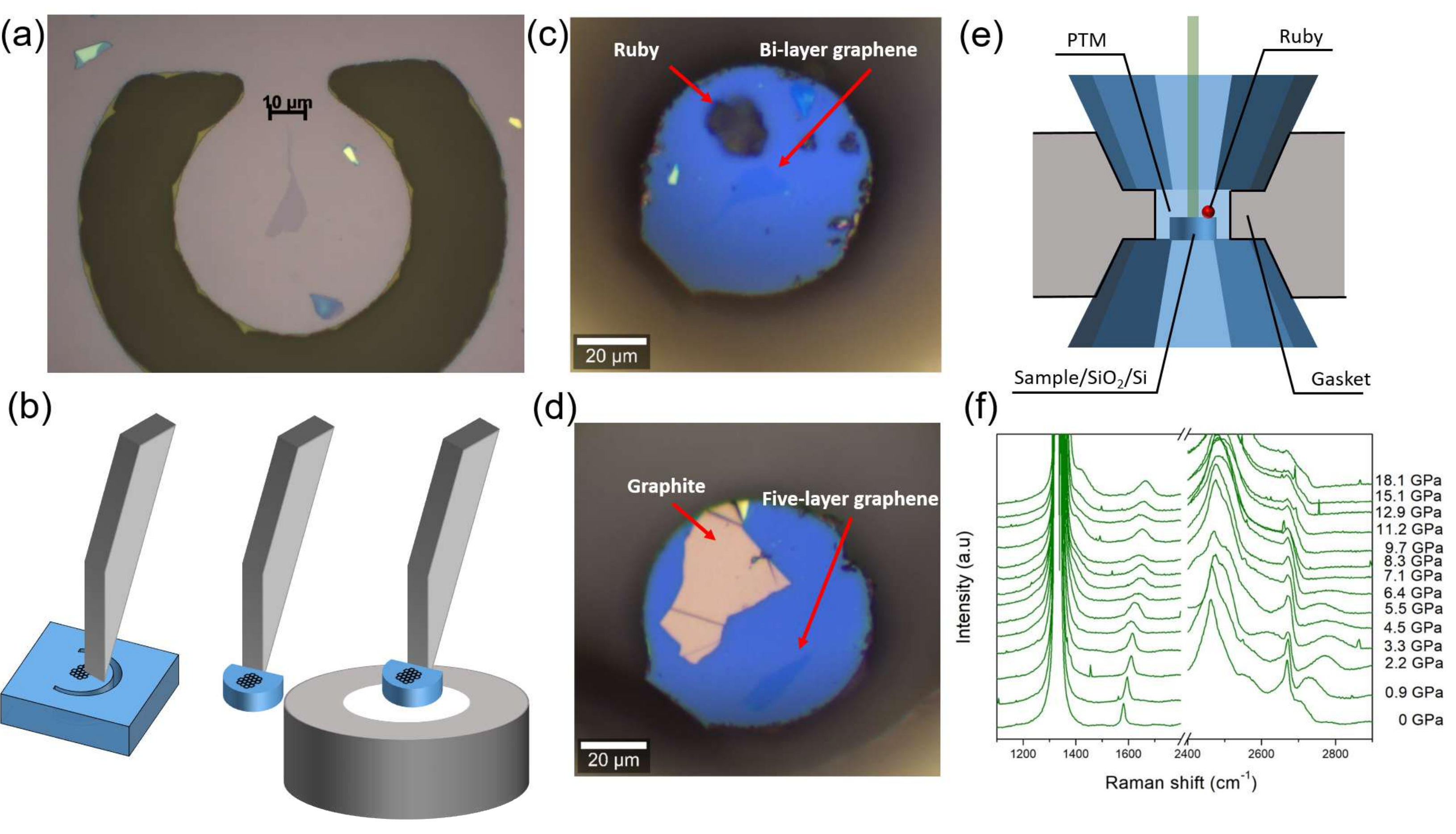}
\caption{\label{fig1} Horseshoe technique to load 2D materials transferred onto SiO$_2$/Si substrates into DACs.(a) Optical image of a bi-layer graphene sample on a silicon disk. A ''horseshoe'' shape is etched through a 25$\mu$m silicon sample coated with a 300nm thermal oxide. Then, the 2D sample is transferred onto the center of the silicon disk via the standard pick up and transfer technique.(b) Illustration of the ''breaking and loading'' process. First, the silicon disk containing the sample is pressed with the tip of a needle, in a region away from the sample, until it cleaves. The sample/SiO$_2/$Si is then picked up, and loaded into the gasket hole with the aid of a needle. (c-d) Optical images of the samples inside the gasket hole of bi-layer graphene (c) and five-layer graphene and graphite (d) The bi-layer graphene sample is the same one shown in (a). (e) Schematic of the High-Pressure Raman experiments. The sample was loaded into a DAC together with the ruby and water PTM. At each pressure, the Raman spectra was acquired with the 532nm laser excitation.(f) Evolution of the G and 2D bands (at 1580 and 2700 cm$^{-1}$ at ambient pressure, respectively) with pressure for 5-layer graphene sample compressed in water PTM up to 18\,GPa. The strong intensity features at approximately 1330 and bellow 2670 cm$^{-1}$ correspond to the first and second-order Raman peaks from the top diamond of the DAC.}
\end{center}
\end{figure}

Horseshoe-shaped trenches were etched through 25-$\mu$m-thick Si substrates covered with a 300-nm-thick SiO$_{2}$ layer as shown in Figure~\ref{fig1}a. The graphene flakes were deposited onto the regions surrounded by the trenches, using the pick--up/transfer technique described in Ref.~\cite{wang2015electronic}. Afterward, the tiny ($\approx$\,70\,$\mu$m of diameter) SiO$_{2}$/Si disk supporting the graphene piece is cleaved and detached from the SiO$_{2}$/Si wafer using a sharp stainless steel tip, and the same tip is used to bring the disk inside the diamond anvil cell (DAC), as illustrated in Fig.~\ref{fig1}b.\\

The graphene powder was prepared by liquid--phase exfoliation of natural graphite. The sample was centrifuged at 15\,kG. Statistical AFM analysis~\cite{Thales2019} was carried out over more than 4000 flakes, and the results show that $\approx\,25\%$ of the sample, in mass, is composed of few--layer graphene flakes (less than 5 layers), $\approx\,35\%$ is composed of graphene platelets with thicknesses between 6 and 10 layers, $\approx\,26\%$ is composed of platelets with thicknesses between 10 and 20 layers, and $\approx\,14\%$ of flakes with more than 20 layers. In fact, from the XRD data taken at ambient pressure, one can notice that the G002 and G101 peaks from graphene are broad, an indication of poor degree of crystallinity along the $c$ axis, as expected for a powder composed of randomly stacked 2D flakes. To avoid the presence of residual water content and reagents, the sample was heated up to 400\,C$^{\rm o}$.\\

\subsection{High-pressure Raman}

Raman spectra were acquired using an alpha 300 system RA from WITec (Ulm, Germany) equipped with a highly linear (0.02\%) piezo-driven stage, and objective lenses from Nikon (10$\times$, NA = 0.3 for high-pressure measurements and 50$\times$, NA=0.55 for room condition measurements). A diode pumped solid state polarized laser, $\lambda$ = 532 nm, was used. The incident laser was focused with a diffraction-limited spot size (0.61$\lambda$/NA), and the Raman signal was detected by a high-sensitivity, back-illuminated CCD located behind a 600 gmm$^{-1}$ grating. The spectrometer used was an ultra-high throughput Witec UHTS 300 with up to 70\% throughput, designed specifically for Raman microscopy.  To avoid damage due to sample heating, measurements were performed with powers of approximately 3\,mW and 9\,mW for the 50$\times$ and 10$\times$ objectives, respectively.\\

High-pressure Raman measurements were done with a gas-membrane driven DAC, microScope DAC-HT(G), using Inconel 718 pre-indented ($\sim 90 \mu$m) gaskets.  It is important to note that at pressures below 3\,GPa, the 2D peak partially overlaps with the second-order Raman peak originated from the DAC's diamond anvils. The pressure was monitored using the ruby's calibration method described in Ref.~\cite{mao1986calibration}. The spectra of G and 2D bands were fitted with Voigt functions using the software PeakFit v4.12. The numbers of layers of graphene samples were determined by standardized principal component analysis (PCA) of the 2D band\cite{Diego2019}.\\

\subsection{Indentation experiments}

The high-pressure compression of the 4-layer graphene and graphite on SiO$_2$ substrate sample was performed with a Symmetric type DAC, using stainless steel pre-indented ($\sim 90 \mu$m) gaskets. The pressure was increased to 8\,GPa and kept at this pressure for 12 hours, after which, the DAC was opened and the sample was recovered and its spectra collected. No Raman spectra were acquired during compression. Raman spectra were acquired after decompression using a confocal microscope spectrometer (Horiba LabRAM HR Evolution) in a backscattering geometry, with a 50$\times$ objective lens and 532 nm laser. The measurement was performed with a power of approximately 3\,mW to avoid damage due to sample heating.\\

The AFM topography measurement of the recovered samples was performed with a Cypher S AFM Microsope using an AC240TS-R3 cantilever from Oxford Instruments, with resonance frequency at 70\,Hz.\\

\subsection{High-pressure x-ray diffraction}

High-pressure XRD measurements were conducted at 16ID-B of High Pressure Collaborative Access Team (HPCAT) at the Advanced Photon Source(APS), Argonne National Laboratory, using a symmetric type DAC. The incident monochromatic x-ray beam, $\lambda = 0.40663$\,\AA\,was focused down to 5-10\,$\mu$m in diameter, and the XRD patterns were collected on a Pilatus Area Detector. The  XRD spectra were analyzed using Matlab for background subtraction and PeakFit v4.12 for Voight fitting the G002 peak.

\subsection{Theoretical calculations details}

In the MD simulations, performed with the LAMMPS~\cite{plimpton1995fast} software, atoms were treated as point particles interacting through the AIREBO~\cite{stuart2000reactive} potential.  The H-functionalized five-layer graphene was represented by 22.880 and 2.288 carbon and hydrogen atoms, respectively, and  boundary conditions were imposed in the in-plane x- and y- directions. The time step was set to 0.25 fs.  We employed the Nos\'e-Hoover thermostat~\cite{nose1984unified} scheme to keep  the average temperature at 300 K. Additionally, the Nos\'e-Hoover barostat was used in order to allow for fluctuations in the lateral (xy) box dimensions (external pressure set to 1 bar).  The five layers were  placed in the AB stacking, with a repulsive wall placed immediately bellow the fifth layer.  The pressure was applied to the system by a moving piston modeled as a  Lennard-Jones (LJ) wall, initially placed 4\,\AA\, above hydrogen atoms, with energy and distance parameters of 1.0 meV and 3.0\,\AA, respectively.  Simulations were divided in two stages, the preparation and production ones. The preparation stage comprised a structure minimization followed by  a temperature equilibration dynamics, in which the system was linearly heated from 30\,K to 300\,K during\,25 ps and kept at this temperature for further 25\,ps. In the first part of the production stage - the loading process -, the distance between the piston and the system was linearly decreased during 200\,ps, leaving an average graphene interlayer separation of 2.6\,\AA. Piston position was kept for 400\,ps, and then returned to its initial position after additional  100\,ps.\\

In the DFT~\cite{hohenberg1964inhomogeneous,kohn1965self} calculations, we employed the SIESTA implementation~\cite{soler2002siesta}, making use of norm conserving Troullier-Martins pseudopotentials~\cite{troullier1991efficient} in the factorized Kleinman-Bylander form~\cite{kleinman1982efficacious}. The basis set was composed of double-zeta pseudoatomic orbitals of finite range augmented by polarization functions - the DZP basis set. The generalized gradient approximation (GGA)~\cite{perdew1996generalized} was chosen to represent the exchange-correlation functional. We performed integrals in real space using a grid defined by a meshcutoff of 350 Ry, and the Brillouin zone was sampled using a k-grid cutoff of 30\,\AA. The geometries were considered optimized when the maximum force component (not constrained) in any atom was less than 10 meV/ \,\AA. To mimic the application of pressure, we employ a hard-wall constraint in the lowest carbon atoms, which are not allowed to move in the negative z--direction, while a predefined force is applied in the oxygen atoms of the --OH groups which functionalize the upper graphene surface.\\

In the MD simulations, we chose hydrogen functionalization of the upper graphene layer since  model potentials for C-H bonds are well described in literature. DFT calculations can be performed with either --OH and --H groups, rendering basically the same results. We chose -OH to be consistent
with similar models reported in previous studies~ \cite{barboza2011room,martins2017raman}. For plotting the LDOS of the dangling bonds in Figure~6a-b, the Xcrysden program was used~\cite{kokalj2003computer}. An isovalue of 0.01 was used in Fig.~6a and 0.015 in Fig.~6b. The XRD simulation of BB bulk diamondene was obtained from Rietveld refinement using the RIETAN-FP software~\cite{izumi2007} as implemented in the VESTA program`\cite{Momma:db5098}.

\subsection{Hydrostatic limit of water PTM}

Piermarini et. al~\cite{piermarini1973hydrostatic} investigated the hydrostatic limit of water, among other PTMs, in DAC experiments using two methods: (I) measuring the pressure at several different locations from ruby crystals spread across the chamber and (II) by measuring the spectral linewidth of the R$_1$ fluorescence peak from ruby.  From (I), they observed an unexpected low-pressure gradient to pressures above 10\,GPa and up to $\sim$ 14\,GPa with practically no pressure gradient up to 10 GPa and from (II), they observed  an increase in linewidth that is consistent with the observations from (I). Olinger et. al~\cite{olinger1975compression} performed high-pressure XRD on a mixture of aluminum powder and distilled water, and they reported no effects due to localized non-hydrostatic stresses or detection of pressure gradients over their pressure range, which was up to 8\,GPa. Our analysis of the linewidth of the R$_1$ and difference in peak position between R$_1$ and R$_2$ ruby fluorescence peaks, shows no indications of non-hydrostatic stress until $\sim$ 10\,GPa and supports those observations from previous works. It is important to mention that most of our evidence of phase transition from the High-Pressure Raman experiments were obtained in a pressure range bellow 7\, GPa.

\subsection{Optical changes of 5-layer graphene and graphite samples during compression}

\begin{figure}[!tb]
\begin{center}
\includegraphics [scale=0.75]{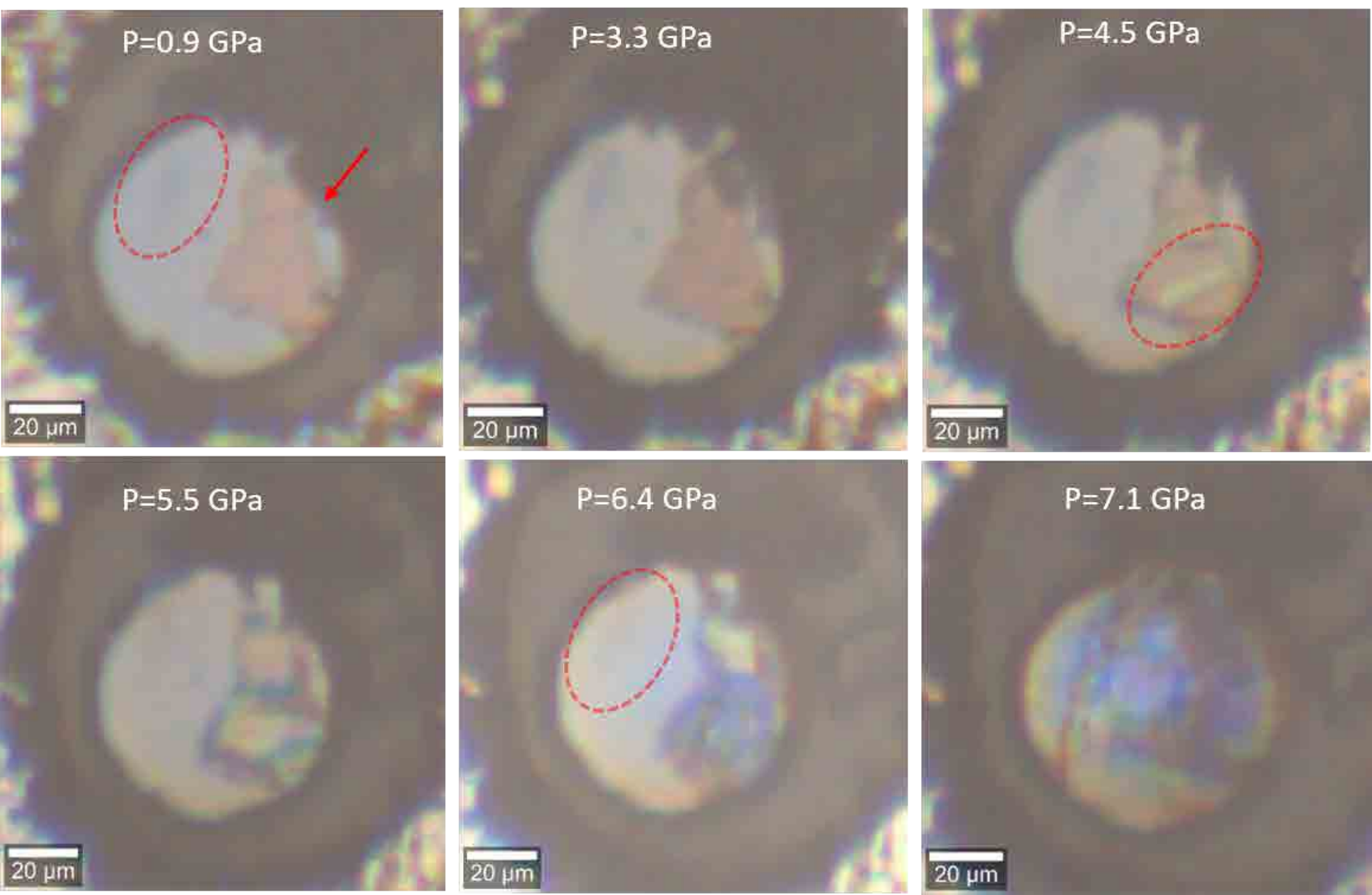}
\caption{\label{fig18} Evidence of a gradual top-bottom transparency for 5-layer graphene and graphite samples during compression. Optical images of 5-layer graphene and graphite samples inside of the DAC at different pressures.In the first panel (0.9\,GPa), the 5-layer graphene is highlighted by the dashed ellipse, and the graphite flake is indicated by the arrow.  At 4.5\,GPa, a color change occurs in the graphite flake, with the formation of a yellow strip. With increasing pressure to 5.5\,GPa, the yellow region spreads over the graphite flake. At 6.4\,GPa, the 5--layer graphene flake turns transparent. The yellow color remains only at the top-half region of the graphite flake, and its bottom-half shows a color similar to the substrate -- an indication of transparency. At 7.1\,GPa, the color changing ends, and the flakes can no longer be seen. The subsequent color changes are likely related to fractures on the substrate, damaging the SiO$_2$ layer.}
\end{center}
\end{figure}

Consistent with the Raman spectroscopic signatures of a pressure-induced phase transition, changes in the optical properties of the 5-layer graphene and graphite flakes were also observed in the 4--7\,GPa range. Figure~\ref{fig18} shows optical images of the sample inside the DAC at different pressures. Starting at 4.5\,GPa, a series of color changes occur in the graphite flake, with the formation of a yellow region which spreads over the flake with increasing pressure, then gradually turns to a color similar to the substrate -- an indication of increasing transparency. The 5--layer graphene and graphite flakes can no longer be seen at 6.4\,GPa and 7.1\,GPa, respectively.\\

Overall, the observation of these changes in the optical properties of the flakes  suggests a gradual top-bottom increase in transparency, consistent with a phase transition starting with the first top layers and propagating to the bottom with increasing pressure.

\subsection{Indentation evidence}  \label{IPTIE}

\begin{figure}[!tb]
\begin{center}
\includegraphics [scale=0.7]{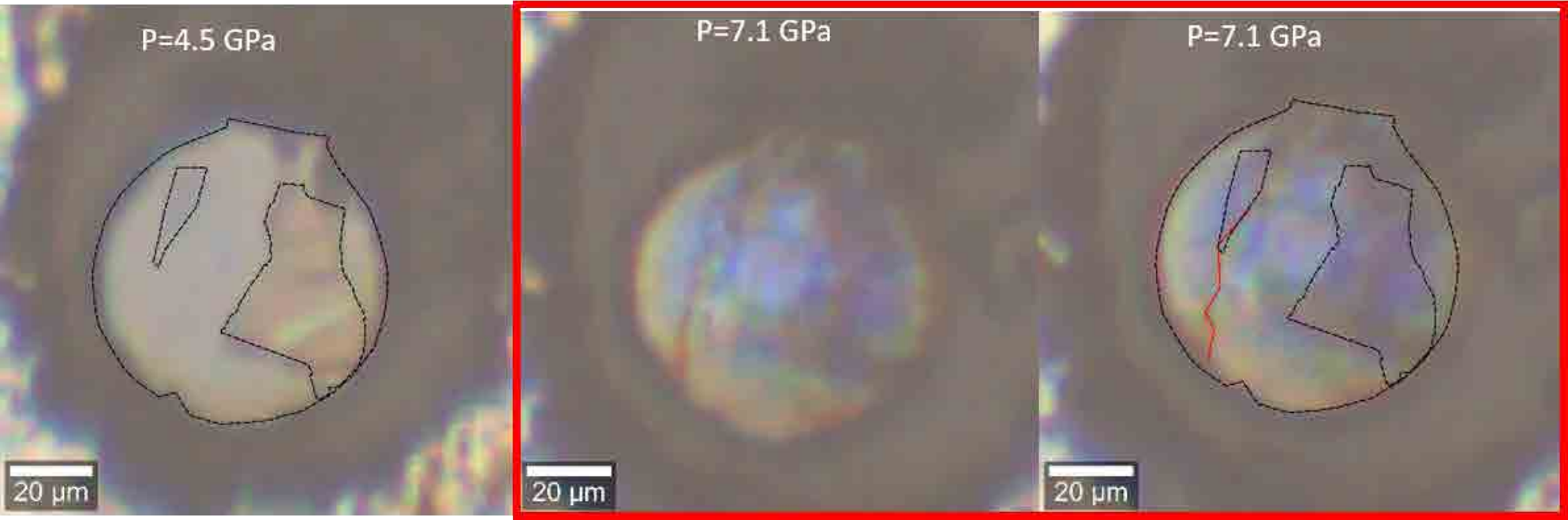}
\caption{\label{fig6b} Evidence for the formation of a hard phase during compression. Optical images highlighting the flake contours by black dashed lines before (left) and after (right, inside red box) phase transition for five-layer graphene. The silicon fractures are highlighted by the red-dashed lines.}
\end{center}
\end{figure}

Figure~\ref{fig6b} shows the formation of cracks on the silicon along directions defined by the edges of the flake after the pressure-induced phase transition of the five-layer sample. Those features are consistent with the formation of a hard-phase upon phase transition.\\

\begin{figure}[!tb]
\begin{center}
\includegraphics [scale=0.55]{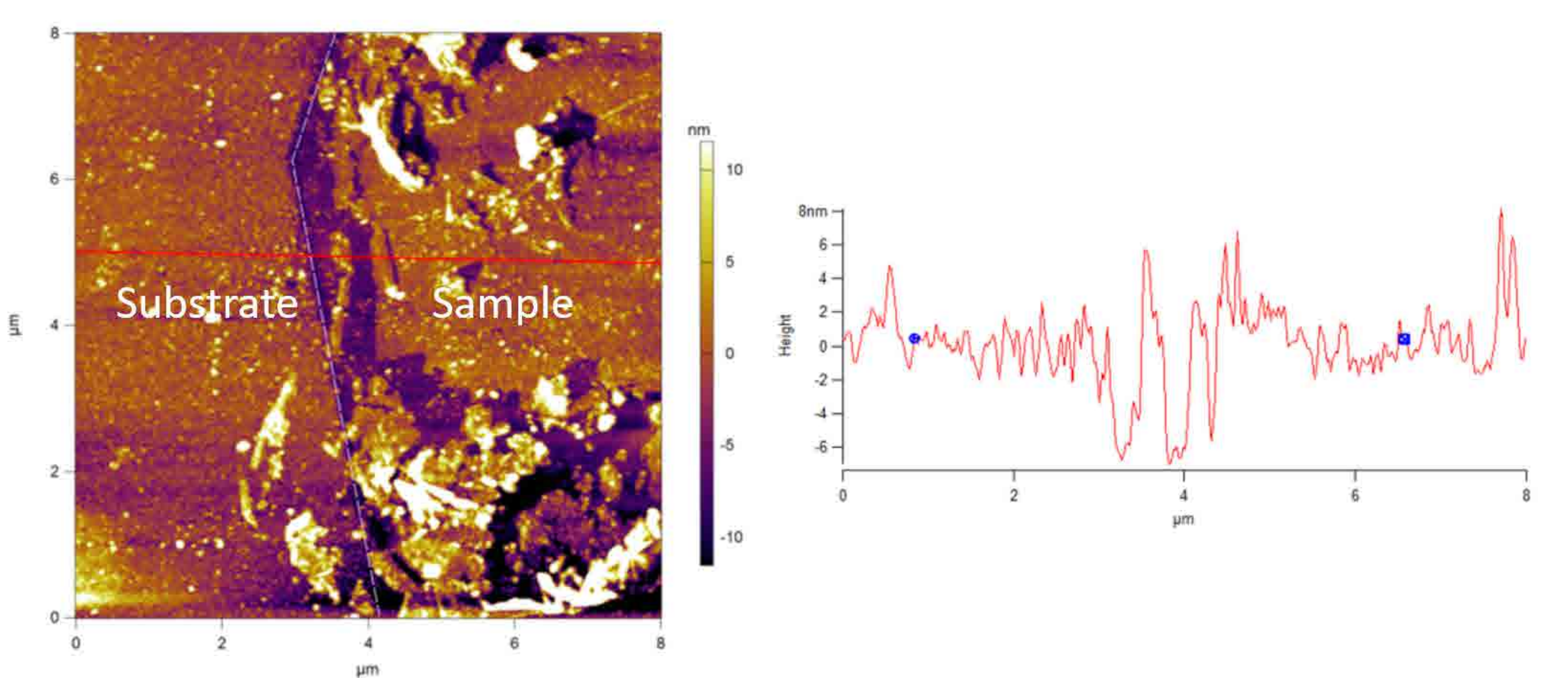}
\caption{\label{fig7b} Evidence of indentation of the SiO$_2$/Si substrate from the recovered four-layer graphene sample. AFM topographical images of the four-layer graphene flake after compression, showing the deep grooves formed on the silicon oxide substrate along the edges of the flake. The white dashed line separates the SiO$_2$ substrate and flake regions.}
\end{center}
\end{figure}

Figure~\ref{fig7b} shows the AFM topographical image of the four-layer sample after compression, where it is possible to see the grooves on the silicon substrate along the edges of the flake. For comparison, indention experiments on SiO$_2$~\cite{cellini2019layer} show that this material can only be indented by a few angstroms in the estimated pressure range of 0-7 GPa. However, the deep grooves formed along the edges of the flakes --as shown by the AFM image -- indicate that SiO$_2$ was indented by a few nanometers in the current case. Indentation experiments performed on graphene of different thicknesses on SiO$_2$, showed that graphene systems (1--3 layers and graphite) are softer than SiO$_2$~\cite{cellini2019layer}, being unable to cause this type of indentation. This confirms the formation of a hard phase during compression, capable of indenting the silicon oxide substrate.

\subsection{Origin of the -OH and -H chemical groups in diamondene}\label{Origin}

\begin{figure}[!tb]
\begin{center}
\includegraphics [scale=0.25]{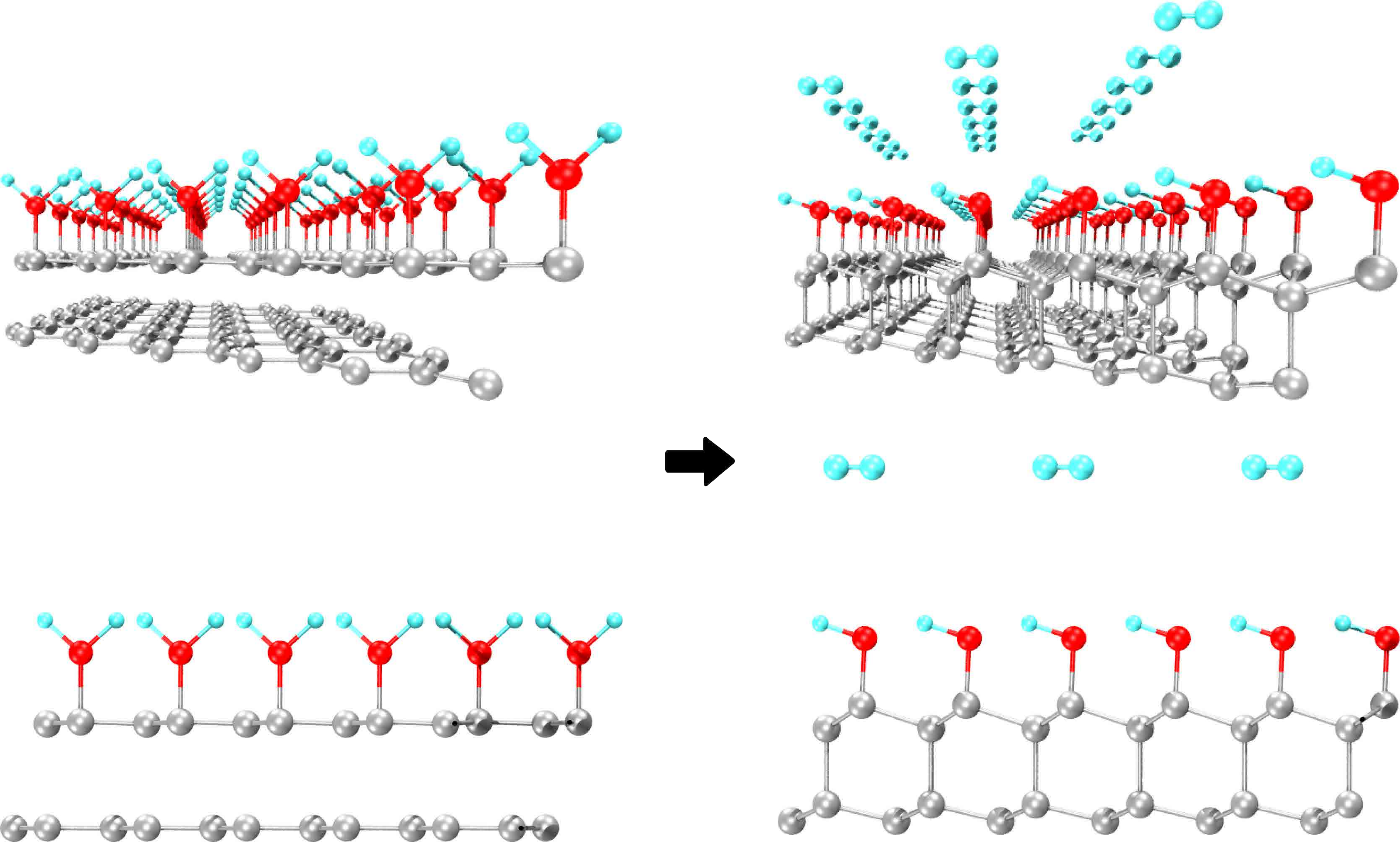}
\caption{\label{fig9} DFT Model for the origin of the -OH and -H chemical groups in diamondene. Two views of initial (left) and final (right) configurations for a graphene bilayer covered with H$_2$O molecules and subjected to pressure. The initial interlayer distance is 2.3\,\AA.}
\end{center}
\end{figure}

The formation of -H or -OH groups in the water-graphene interface under pressure is a complex issue due to the difficulties in simulating a chemical reaction through first principles calculations, and it is beyond the scope of our work. Nevertheless, simplified mechanisms may be proposed to shed light upon the role of the dissociation of water molecules in the rehybridization process. To this purpose, we propose an initial configuration for the first principles geometric optimization in which H$_2$O molecules are placed on top of a bilayer graphene. Then we apply pressure on the system by decreasing the graphene interlayer distance and by constraining the vertical displacements of the bottommost carbon atoms and of the oxygen atoms, the so-called hard-wall constraints. Figure~ \ref{fig9} on the left shows this initial geometry with a interlayer distance of 2.3\,\AA. Upon optimization, a OH bond increases while the graphene layers approach each other becoming increasingly corrugated. Eventually, the O-H bond breaks and the hydrogen atom binds to the neighboring H atom, forming a H$_2$ molecule which becomes physically adsorbed on top of the diamondene structure just formed. This last configuration is shown in Fig.~\ref{fig9}, on the right.

\subsection{Possibility of different rehybridized geometries from DFT calculations}

Due to the importance of kinetic aspects, our DFT calculations do not aim to find the lowest energy configuration. Rather, given the initial conditions (interaction between graphene and ---OH groups), our conclusion is  that, in fact,  mixed stacking orders are highly likely. In a systematic investigation of this problem, Xie et al.~\cite{xie2017graphite} have ascribed this hexagonal diamond order to a facile initial nucleation mechanism, involving the existence of coherent sp$^{2}$--sp$^{3}$ interfaces. Still according to their results, cubic diamond growth should have a much slower kinetics and should be mixed with that of hexagonal diamond.\\

\begin{figure}[!tb]
\begin{center}
\includegraphics [scale=0.75]{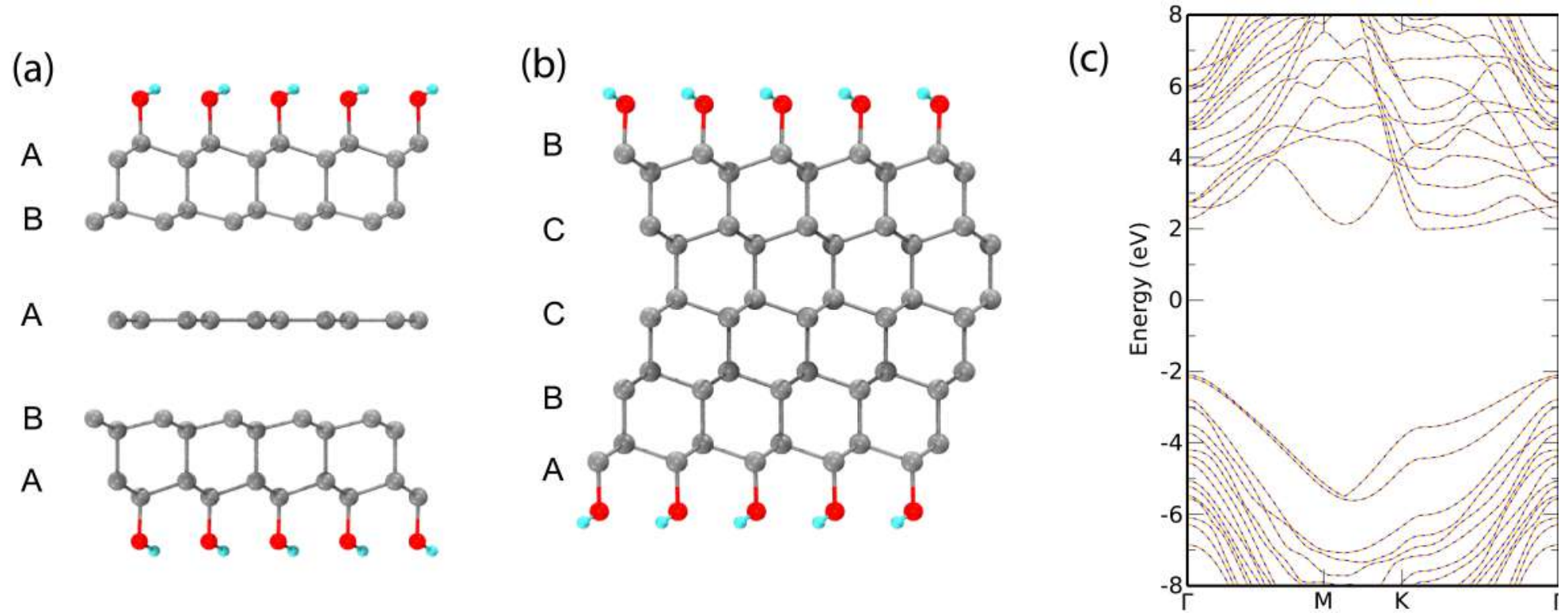}
\caption{\label{fig14} Evolution of the sp$^{2}$--sp$^{3}$ restructuring process considering functionalization on both surfaces. (a--b), DFT results for five-layer system functionalized on both surfaces for two values of applied force. The final system is a mixture of AB and BB stackings.(c), Band structure for the final configuration. The band gap opens up to 4.07 eV.}
\end{center}
\end{figure}

The AB and AA stacking types still coexist in a model which takes into account functionalization on both upper and lower surfaces. In this model, the first rehybridized geometry is symmetric and composed of a pair of diamondene structures separated by a graphene layer, as shown in Fig.~\ref{fig14}a. With increasing pressure, this layer also becomes sp$^{3}$--hybridized, leading to the structure shown in Fig.~\ref{fig14}b -- the resulting stacking is ABCCB. The double functionalization saturates all dangling bonds -- as a result, no localized states appear in the spectrum, and the band gap opens up to 4.07\,eV, as shown if Fig.~\ref{fig14}c.\\

\subsection{Detailing the sp$^{3}$ bond formation from MD simulations}

We further detail the sp$^{2}$-sp$^{3}$ restructuring process of the compressed AB-stacked five-layer graphene. As explained in the main text, the phase transition starts with the diamondization of the first two layers, giving rise to diamondene. This happens in two steps, the first being nucleation: the formation of random chemical bonds between the two layers due to their closer proximity, allowing for the sp$^{2}$--sp$^{3}$ rehybridization of the carbon atoms. Such a process is greatly facilitated by the presence of --H groups on the top surface. Once these chemical bonds are formed, they trigger the sp$^{3}$ conversion across the first two layers through a cooperative phenomenon\cite{stojkovic2003collective} in which each tetrahedral bond favors the formation of a neighboring sp$^{3}$ bond. As a result, the diamondene structure is formed in this so-called horizontal propagation. For the vertical propagation to occur, the pressure needs to be further increased.\\

\begin{figure} 
   \centering
    \includegraphics[scale=0.61]{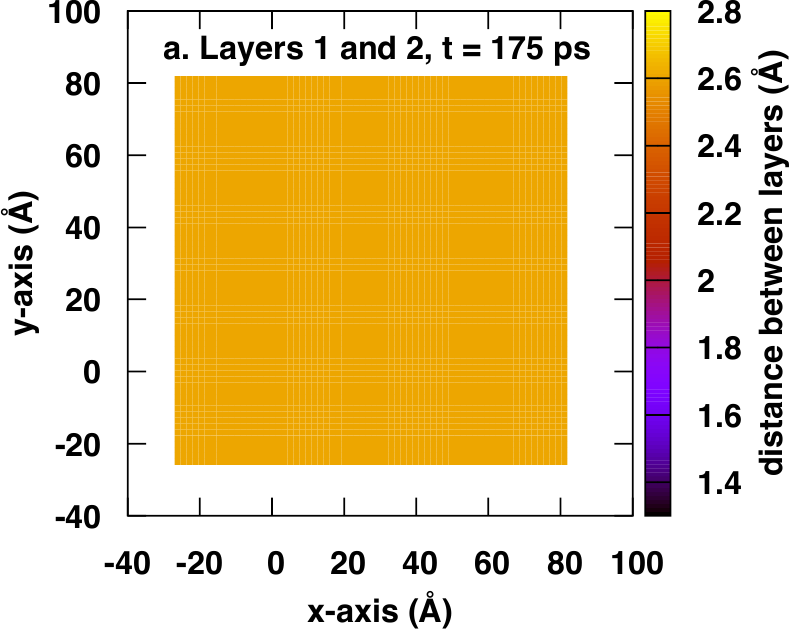}
    \includegraphics[scale=0.61]{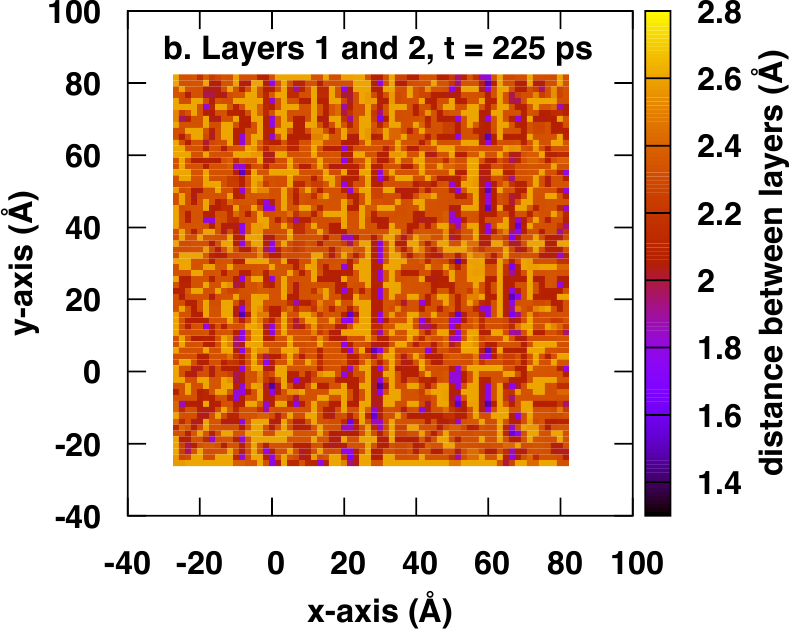}
    \includegraphics[scale=0.61]{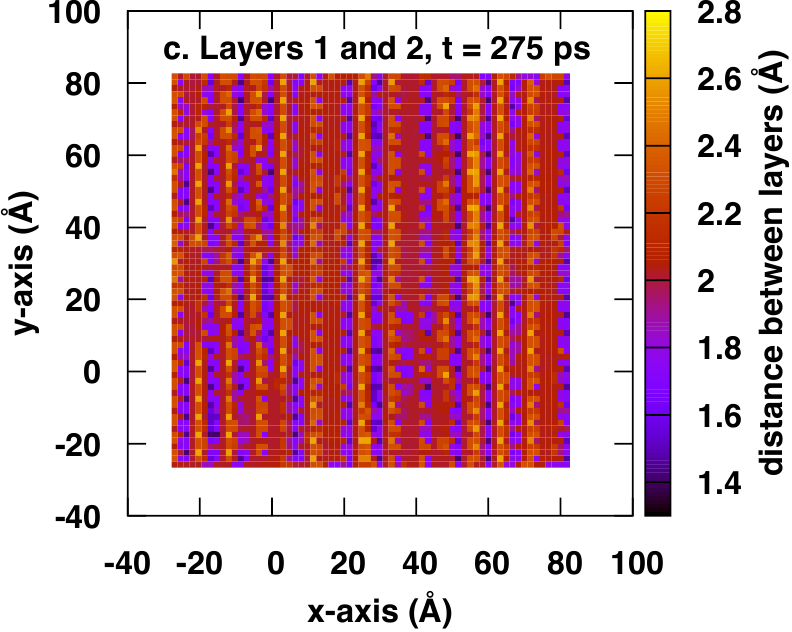}
    \includegraphics[scale=0.61]{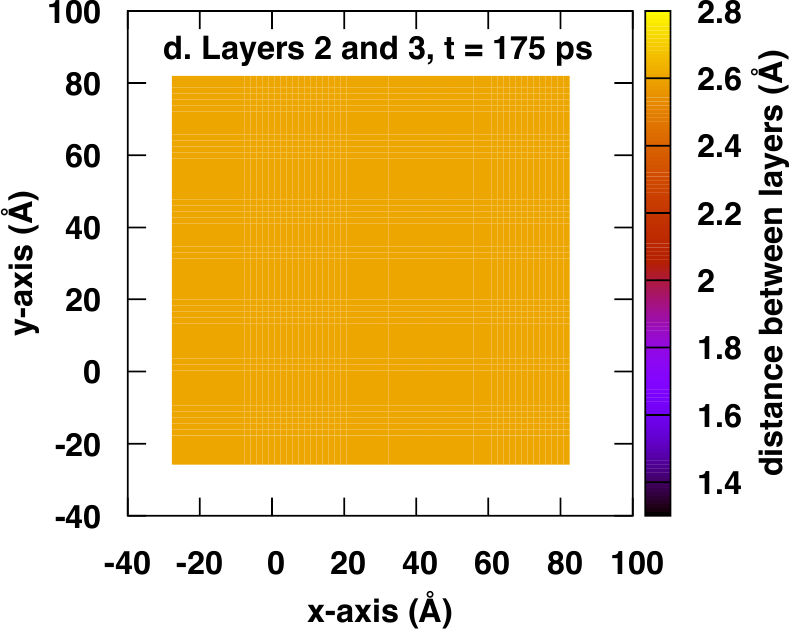}
    \includegraphics[scale=0.61]{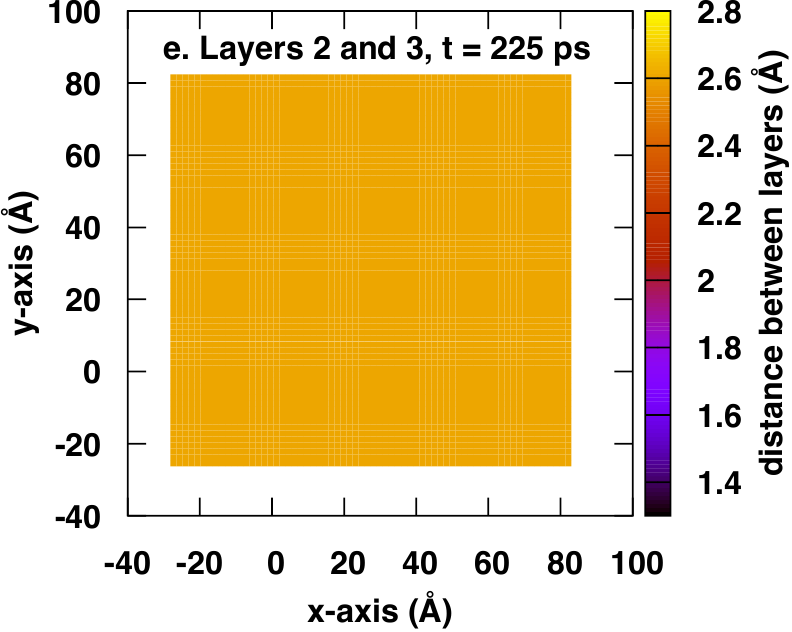}
    \includegraphics[scale=0.61]{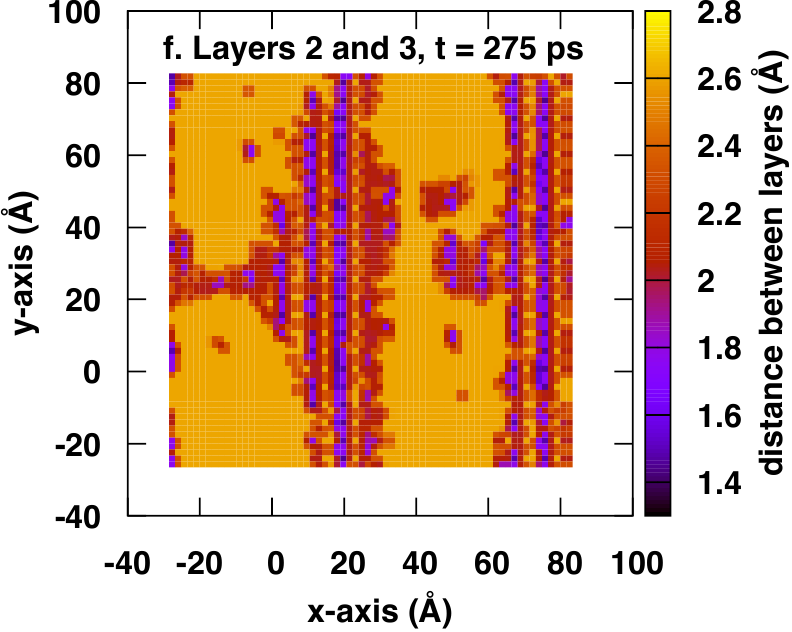}
    \includegraphics[scale=0.61]{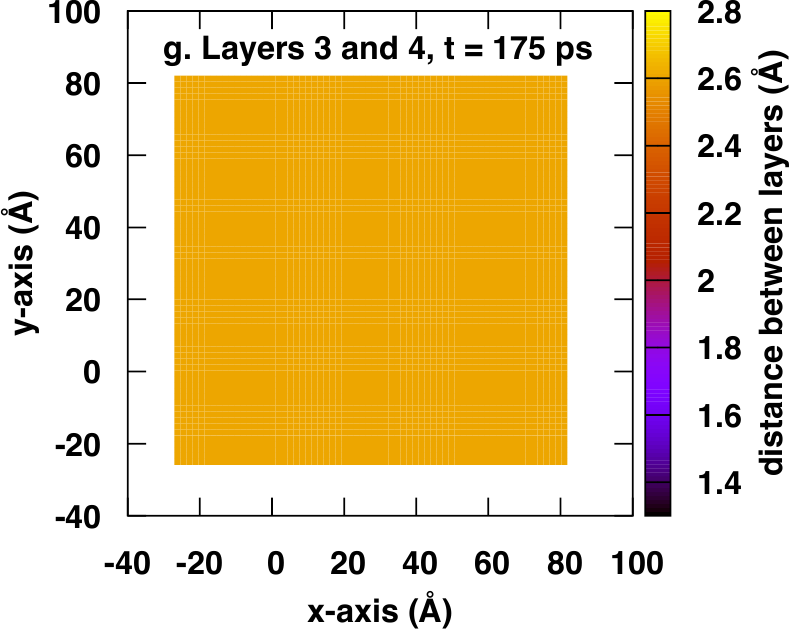}
    \includegraphics[scale=0.61]{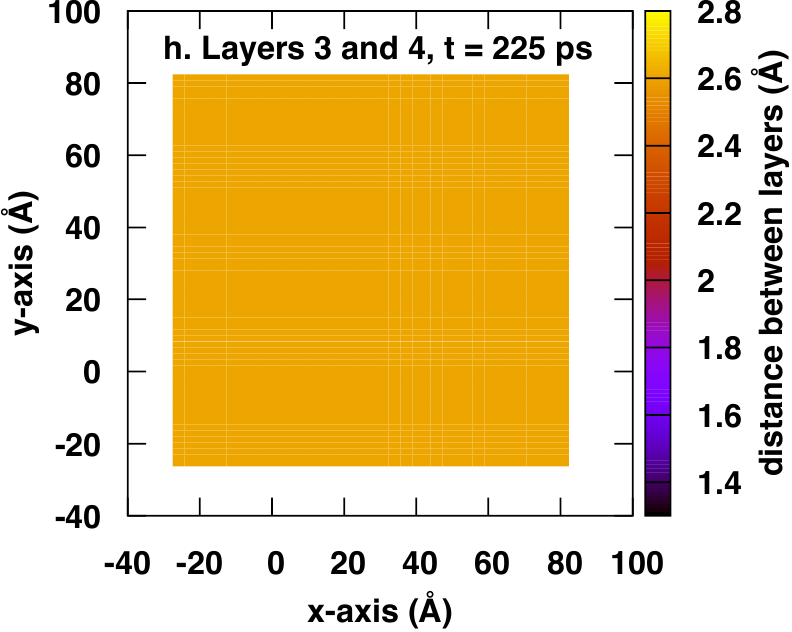}
    \includegraphics[scale=0.61]{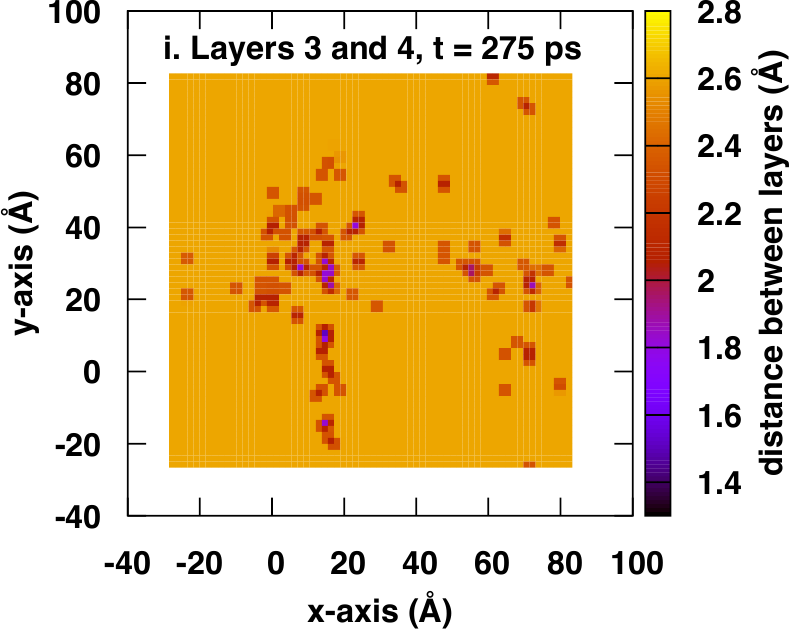}

   \caption{ \label{fig13} Heatmaps showing interlayer distances between pairs of layers during compression for AB-stacked five-layer graphene. Heatmaps showing  distances between (a)-(c) layers 1 and 2,  (d)-(f) layers 3 and 4, and (g)-(i)  layers 3 and 4 for different times. Distances between layers 4-5 are not shown since no sp3 bonds were formed.}
\end{figure}

Figure~\ref{fig13} presents heatmaps in which the colors represent local interlayer distances. Panels show layer pairs distances at three different times: 175 ps, 225 ps, and 275 ps between layers 1 and 2 (a)-(c), 2 and 3 (d)-(f), 3 and 4 (g)-(i). We consider a sp$^{3}$ bond to be formed if the interlayer distance is equal to or less than 1.6\,\AA. Regions in orange have typical interlayer distances of about 2.6\,\AA, which indicates non-sp$^{3}$ bond formation. Regions approaching the purple color, corresponding to a distance of 1.6\,\AA, indicate the sp$^{2}$--sp$^{3}$ phase transition occurrence.\\

From Fig.~\ref{fig13} we see that all possible sp$^{3}$ bonds between layers 1 and 2 are formed, considering the range of pressures of our simulations. Layers
2 and 3 form 45\% of possible sp$^{3}$ bonds while layers 3 and 4 form only 5\%. It is worth mentioning that such sp$^{2}$--sp$^{3}$ transitions occur in a narrow time window of few picoseconds for layers 1 and 2 and of femtoseconds between other pairs.\\

We also see from Fig.~\ref{fig13} the occurrence of sp$^{3}$ bonds across layers in the xy plane. The observation of the vertical blue lines in the second panel of heatmap (t\,=\,275\,ps and bonds between first two layers) is a clear signature of the collective process.

\subsection{High-pressure Raman experiment to further investigate our model}

To further investigate the importance of the stacking order in the graphene system for the propagation processes, we carried out additional high-pressure Raman experiments, compressing a graphite piece and a graphene powder in a DAC using water as the PTM. All Raman spectra were acquired using a 532--nm excitation laser. The graphene powder had the same flake-thickness distribution as the powder used in the High-pressure XRD experiments. The graphite piece and the graphene powder were compressed to a maximum pressure of 8 GPa and 30 GPa, respectively.\\

\begin{figure}[!tb]
\begin{center}
\includegraphics [scale=0.7]{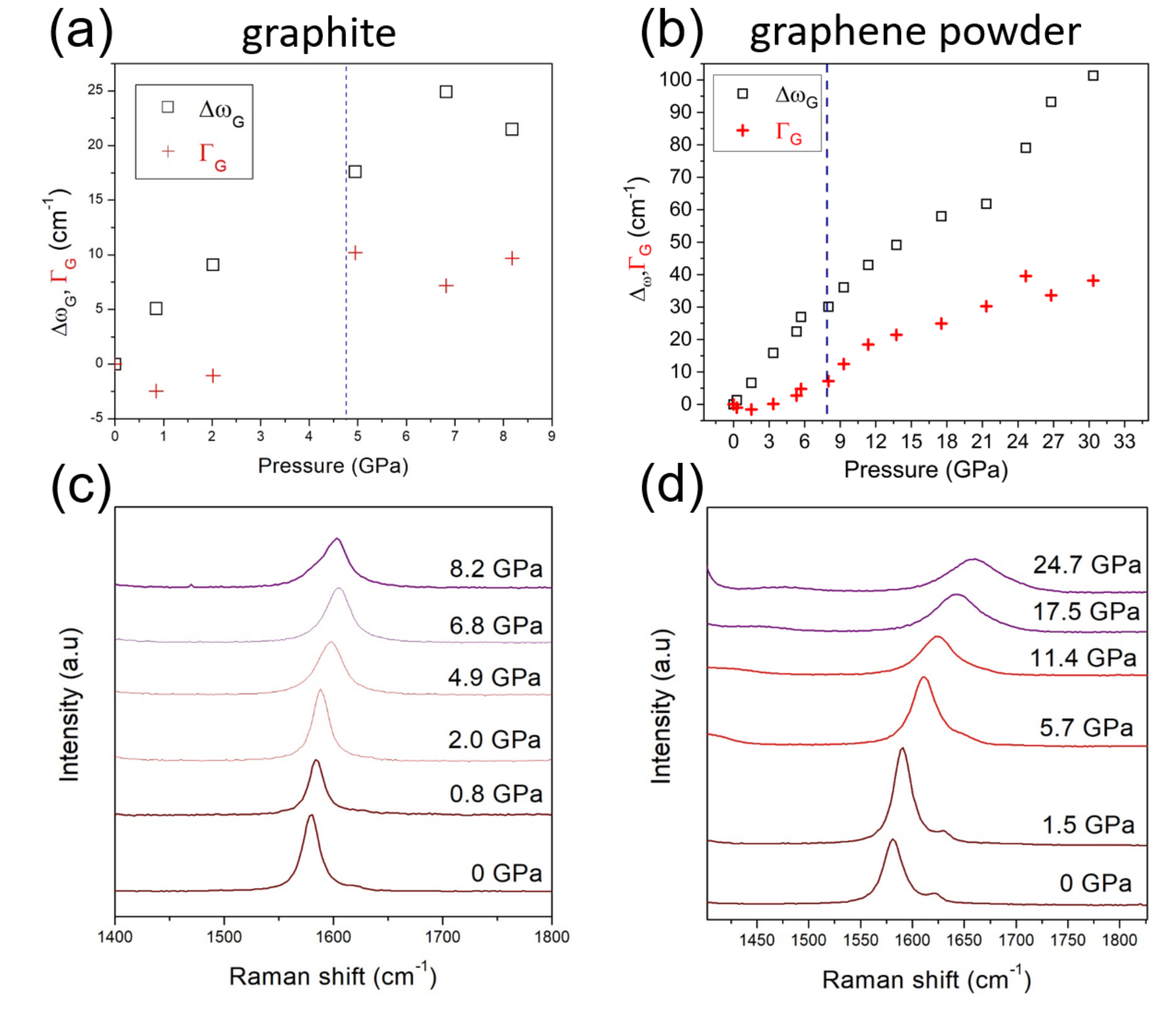}
\caption{\label{fig15} Comparison between Raman spectra features from graphite and a graphene powder under compression. Plots of G band frequency and full width at half maximum  subtracted from their values at initial pressure, $\Delta \omega_G$ and $\Delta \Gamma_G$, respectively, as a function of pressure for (a) graphite and (b) graphene powder. Raman sepctra showing the G band at different pressures for (c) graphite and (d) graphene powder.  }
\end{center}
\end{figure}

Figure.~\ref{fig15}a--b shows a plot of the G band frequency, $\omega_G$, and width $\Gamma_G$ subtracted from their initial values at ambient pressure for graphite and the graphene powder, respectively. In agreement with our high-pressure Raman experiments, both samples showed an abrupt $\Gamma_G$ broadening in the 4--7 GPa range. However, the shape of the G band with increasing pressure for each sample, shown in Figs.~\ref{fig15}c--d for graphite and the graphene powder, respectively, is quite different. After the onset of the phase transition, occurring at $\sim$ 5\,GPa for the graphite piece, and at $\sim$ 7\,GPa for the graphene powder, the G band assumes an asymmetric shape for the former while it remains mostly symmetrical for the latter. Since the main difference between these samples is in their stacking order, the explanation on the difference in G band's shape upon compression, must be related to this feature.\\

The G band asymmetry in graphite was already explained in the main text in terms of scattered Raman signals coming from the coexisting pristine and rehybridized layers during vertical propagation, as illustrated in Fig.~6(a) of the main text. For the graphene powder, due to the random stacking between flakes of different thicknesses, the vertical propagation is inefficient, and one should expect the surface of the powder to consist of rehybridized flakes while the bulk remains unhybridized. Therefore, the Raman spectra of the compressed graphene powder should be largely dominated by the unhybridized flakes from the bulk, giving rise to the symmetrical G band shape. Furthermore, we don't see the clear kinks in $\omega_G \times P$ plots at the critical pressures as observed in Figs.~1(c-d) of the main text (the kink at $\sim$ 24\,GPa is likely related to changes in the compression regime due to non-hydrostatic stress components from the water PTM).\\

\begin{figure}[!tb]
\begin{center}
\includegraphics [scale=0.7]{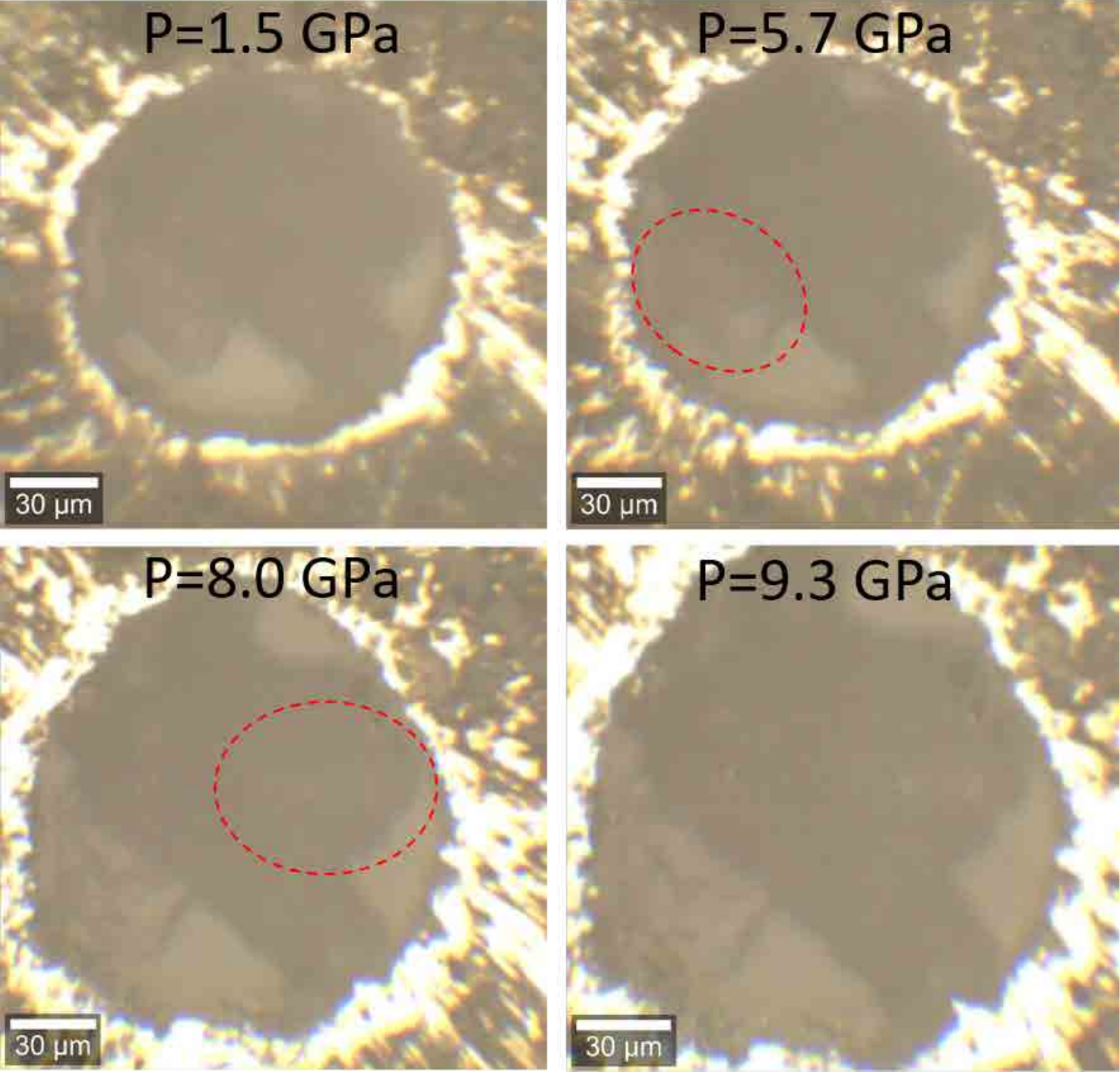}
\caption{\label{fig16} Optical images of the graphene powder inside the DAC at different pressures. The dashed ellipses at 5.7 GPa and 8.0 GPa shows regions of increasing transparency as the pressure is increased.}
\end{center}
\end{figure}

As the pressure increases, the diamondization propagates from the surface to the bulk at a slower rate, when compared to flakes with a defined stacking order. For instance, the graphene powder does not become entirely transparent with increasing pressure, as can be seen from Fig.~\ref{fig16}, which shows optical images of the graphene powder inside the DAC at different pressures. Only certain regions of the powder, shown by the dashed ellipses in Fig.~\ref{fig16}, become visually more transparent. Those regions are probably where the stacking order is more favorable for the vertical propagation to occur.

\subsection{Extinction of G002 peak and pressure range of phase transition}\label{Exctintion}

\begin{figure}[!tb]
\begin{center}
\includegraphics [scale=0.7]{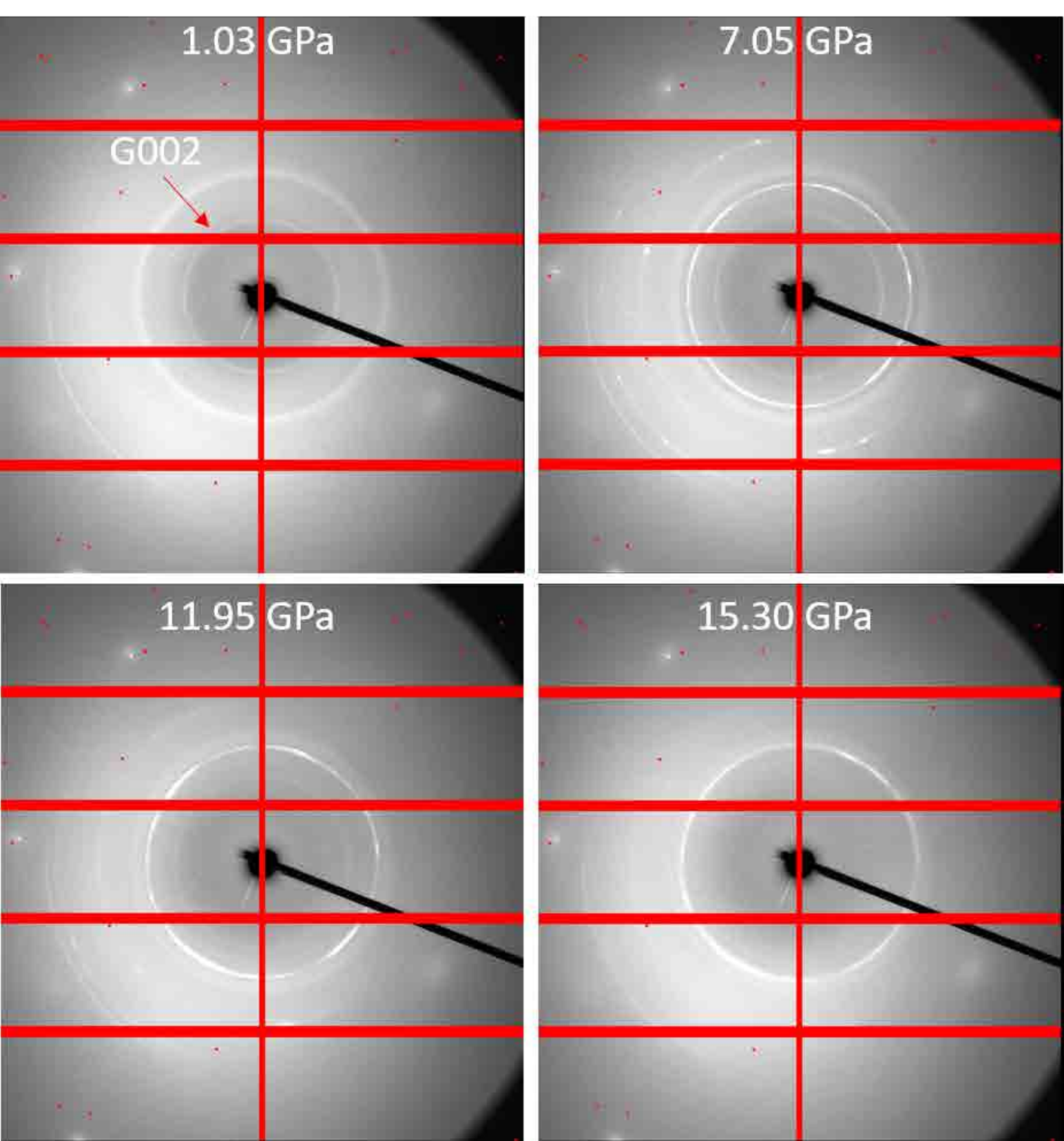}
\caption{\label{fig10} Powder-diffraction images obtained from the Area Detector at different pressures. The red arrow on the top left panel indicates diffraction coming from the G002 peak. }
\end{center}
\end{figure}

It is unlikely that the decrease of G002 peak intensity could be due to a flake alignment to the compression axis\cite{yagi1992high}, since the 002 diffraction pattern remains a ring of relatively uniform intensity through the whole experiment (Fig.~\ref{fig10}), and the PTM does not show non-hydrostatic components until approximately 10\,GPa. Moreover, the large number of flakes and their low aspect ratio (average lateral size of $\approx$\,100\,nm, as revealed by AFM), should avoid them to assume preferred orientation during compression.\\

It is important to note that the long range in pressure of this phase transition, compared to the Raman measurements, can be explained by our model. The random stacking of the flakes in the powder should make the vertical propagation inefficient, as previously explained, requiring more pressure for it to happen and therefore extending the pressure range of the phase transition, compared to the single crystals with well defined stacking order.

\begin{figure}[!tb]
\begin{center}
\includegraphics [scale=0.27]{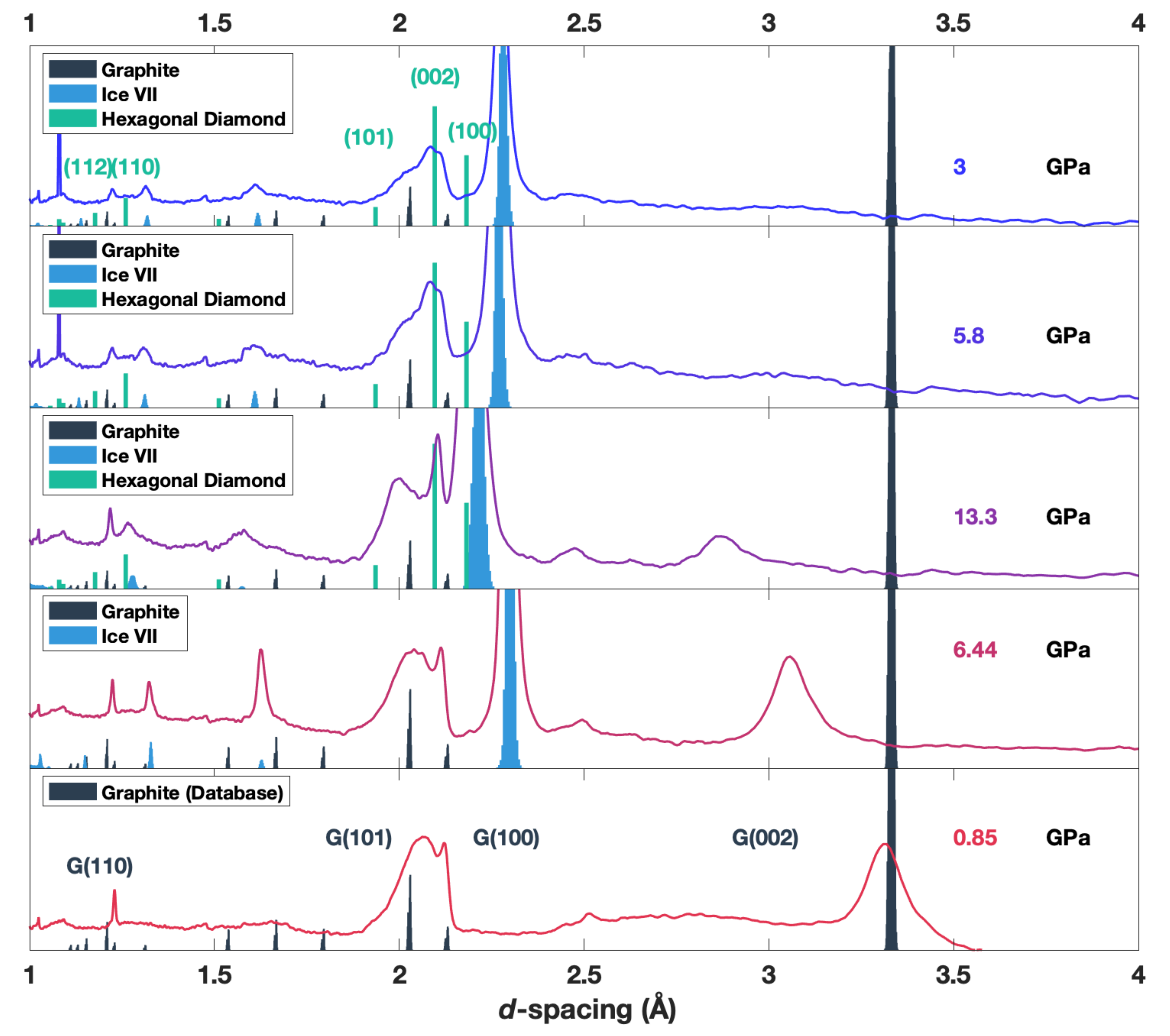}
\caption{\label{fig17} Selected plots of XRD intensity as a function of d-spacing at different pressures for the graphene powder compressed in a water PTM. The graphene powder compressed up to (up arrows) 18\,GPa and decompressed (down arrows) to a residual pressure of 3\,GPa. In the panels, the XRD diffractogram of graphene, Ice VII and AA bulk diamondene/hexagonal diamond are exhibited for comparison. The XRD diffractogram of graphene is obtained at 0 GPa in all panels. The XRD diffractogram of Ice VII were extracted from a separated experiment compressing water up to 11 GPa and, in each panel, the spectra obtained at a pressure closest to the one represented in the panel was chosen. }
\end{center}
\end{figure}

\clearpage

\begin{acknowledgments}
We are grateful to P. Jarillo-Herrero for granting lab access for sample fabrication, and J. Deng, L. Xie, A.P.M. Barboza, L. Dresselhaus-Copper and L. Ju for fruitful discussions. LGPM and JK acknowledge the support from National Science Foundation (NSF) under the EFRI2-DARE program (EFMA-1542863), and from CNPQ under the program Ci\^{e}ncia sem Fronteiras (206251/2014-9). LGPM, MH, TP and JK acknowledge the support from AFOSR FATE MURI, grant no. FA9550-15-1-0514. RC acknowledges support from the Alfred P. Sloan Foundation. LGC and DLS acknowledge the support from CODEMGE. MSCM and ACRS acknowledge financial support from CAPES. LGC, MSCM, ACRS, MJSM, ABO and RJCB acknowledge the support from CNPq and FAPEMIG. ABO, ACRS, LGC, MJSM, MSCM and RJCB acknowledge the support from INCT-Nanocarbono. MJSM and RJCB acknowledge the support from UFOP. This work was performed in part at CNS, a member of NNCI, which is supported by NSF, award no. 1541959. Synchrotron X-ray diffraction was performed at HPCAT (Sector 16), Advanced Photon Source (APS), Argonne National Laboratory. HPCAT operations are supported by DOE-NNSA's Office of Experimental Sciences.  The Advanced Photon Source is a U.S. DOE Office of Science User Facility operated for the DOE Office of Science by Argonne National Laboratory under Contract No. DE-AC02-06CH11357. The high-pressure Raman measurements were performed at LCPNano -- UFMG. The authors acknowledge the MGgraphene Project for providing graphene powder samples and related AFM data.
\end{acknowledgments}


\begin{thebibliography}{46}%
\makeatletter
\providecommand \@ifxundefined [1]{%
 \@ifx{#1\undefined}
}%
\providecommand \@ifnum [1]{%
 \ifnum #1\expandafter \@firstoftwo
 \else \expandafter \@secondoftwo
 \fi
}%
\providecommand \@ifx [1]{%
 \ifx #1\expandafter \@firstoftwo
 \else \expandafter \@secondoftwo
 \fi
}%
\providecommand \natexlab [1]{#1}%
\providecommand \enquote  [1]{``#1''}%
\providecommand \bibnamefont  [1]{#1}%
\providecommand \bibfnamefont [1]{#1}%
\providecommand \citenamefont [1]{#1}%
\providecommand \href@noop [0]{\@secondoftwo}%
\providecommand \href [0]{\begingroup \@sanitize@url \@href}%
\providecommand \@href[1]{\@@startlink{#1}\@@href}%
\providecommand \@@href[1]{\endgroup#1\@@endlink}%
\providecommand \@sanitize@url [0]{\catcode `\\12\catcode `\$12\catcode
  `\&12\catcode `\#12\catcode `\^12\catcode `\_12\catcode `\%12\relax}%
\providecommand \@@startlink[1]{}%
\providecommand \@@endlink[0]{}%
\providecommand \url  [0]{\begingroup\@sanitize@url \@url }%
\providecommand \@url [1]{\endgroup\@href {#1}{\urlprefix }}%
\providecommand \urlprefix  [0]{URL }%
\providecommand \Eprint [0]{\href }%
\providecommand \doibase [0]{https://doi.org/}%
\providecommand \selectlanguage [0]{\@gobble}%
\providecommand \bibinfo  [0]{\@secondoftwo}%
\providecommand \bibfield  [0]{\@secondoftwo}%
\providecommand \translation [1]{[#1]}%
\providecommand \BibitemOpen [0]{}%
\providecommand \bibitemStop [0]{}%
\providecommand \bibitemNoStop [0]{.\EOS\space}%
\providecommand \EOS [0]{\spacefactor3000\relax}%
\providecommand \BibitemShut  [1]{\csname bibitem#1\endcsname}%
\let\auto@bib@innerbib\@empty
\bibitem [{\citenamefont {Zhao}\ \emph {et~al.}(2016)\citenamefont {Zhao},
  \citenamefont {Xu},\ and\ \citenamefont {Tian}}]{zhao2016recent}%
  \BibitemOpen
  \bibfield  {author} {\bibinfo {author} {\bibfnamefont {Z.}~\bibnamefont
  {Zhao}}, \bibinfo {author} {\bibfnamefont {B.}~\bibnamefont {Xu}},\ and\
  \bibinfo {author} {\bibfnamefont {Y.}~\bibnamefont {Tian}},\ }\bibfield
  {title} {\bibinfo {title} {Recent advances in superhard materials},\
  }\href@noop {} {\bibfield  {journal} {\bibinfo  {journal} {Ann. Rev. Mat.
  Res.}\ }\textbf {\bibinfo {volume} {46}},\ \bibinfo {pages} {383} (\bibinfo
  {year} {2016})}\BibitemShut {NoStop}%
\bibitem [{\citenamefont {Balandin}(2011)}]{balandin2011thermal}%
  \BibitemOpen
  \bibfield  {author} {\bibinfo {author} {\bibfnamefont {A.~A.}\ \bibnamefont
  {Balandin}},\ }\bibfield  {title} {\bibinfo {title} {Thermal properties of
  graphene and nanostructured carbon materials},\ }\href@noop {} {\bibfield
  {journal} {\bibinfo  {journal} {Nat. Mat.}\ }\textbf {\bibinfo {volume}
  {10}},\ \bibinfo {pages} {569} (\bibinfo {year} {2011})}\BibitemShut
  {NoStop}%
\bibitem [{\citenamefont {Chernozatonskii}\ \emph {et~al.}(2009)\citenamefont
  {Chernozatonskii}, \citenamefont {Sorokin}, \citenamefont {Kvashnin},\ and\
  \citenamefont {Kvashnin}}]{chernozatonskii2009diamond}%
  \BibitemOpen
  \bibfield  {author} {\bibinfo {author} {\bibfnamefont {L.~A.}\ \bibnamefont
  {Chernozatonskii}}, \bibinfo {author} {\bibfnamefont {P.~B.}\ \bibnamefont
  {Sorokin}}, \bibinfo {author} {\bibfnamefont {A.~G.}\ \bibnamefont
  {Kvashnin}},\ and\ \bibinfo {author} {\bibfnamefont {D.~G.}\ \bibnamefont
  {Kvashnin}},\ }\bibfield  {title} {\bibinfo {title} {Diamond-like c 2 h
  nanolayer, diamane: Simulation of the structure and properties},\ }\href@noop
  {} {\bibfield  {journal} {\bibinfo  {journal} {J. Exp. Theor. Phys.}\
  }\textbf {\bibinfo {volume} {90}},\ \bibinfo {pages} {134} (\bibinfo {year}
  {2009})}\BibitemShut {NoStop}%
\bibitem [{\citenamefont {Chernozatonskii}\ \emph {et~al.}(2010)\citenamefont
  {Chernozatonskii}, \citenamefont {Sorokin}, \citenamefont {Kuzubov},
  \citenamefont {Sorokin}, \citenamefont {Kvashnin}, \citenamefont {Kvashnin},
  \citenamefont {Avramov},\ and\ \citenamefont
  {Yakobson}}]{chernozatonskii2010influence}%
  \BibitemOpen
  \bibfield  {author} {\bibinfo {author} {\bibfnamefont {L.~A.}\ \bibnamefont
  {Chernozatonskii}}, \bibinfo {author} {\bibfnamefont {P.~B.}\ \bibnamefont
  {Sorokin}}, \bibinfo {author} {\bibfnamefont {A.~A.}\ \bibnamefont
  {Kuzubov}}, \bibinfo {author} {\bibfnamefont {B.~P.}\ \bibnamefont
  {Sorokin}}, \bibinfo {author} {\bibfnamefont {A.~G.}\ \bibnamefont
  {Kvashnin}}, \bibinfo {author} {\bibfnamefont {D.~G.}\ \bibnamefont
  {Kvashnin}}, \bibinfo {author} {\bibfnamefont {P.~V.}\ \bibnamefont
  {Avramov}},\ and\ \bibinfo {author} {\bibfnamefont {B.~I.}\ \bibnamefont
  {Yakobson}},\ }\bibfield  {title} {\bibinfo {title} {Influence of size effect
  on the electronic and elastic properties of diamond films with nanometer
  thickness},\ }\href@noop {} {\bibfield  {journal} {\bibinfo  {journal} {J.
  Phys. Chem. C}\ }\textbf {\bibinfo {volume} {115}},\ \bibinfo {pages} {132}
  (\bibinfo {year} {2010})}\BibitemShut {NoStop}%
\bibitem [{\citenamefont {Barboza}\ \emph {et~al.}(2011)\citenamefont
  {Barboza}, \citenamefont {Guimar\~aes}, \citenamefont {Massote},
  \citenamefont {Campos}, \citenamefont {Barbosa~Neto}, \citenamefont
  {Can\c{c}ado}, \citenamefont {Lacerda}, \citenamefont {Chacham},
  \citenamefont {Mazzoni},\ and\ \citenamefont {Neves}}]{barboza2011room}%
  \BibitemOpen
  \bibfield  {author} {\bibinfo {author} {\bibfnamefont {A.~P.}\ \bibnamefont
  {Barboza}}, \bibinfo {author} {\bibfnamefont {M.~H.}\ \bibnamefont
  {Guimar\~aes}}, \bibinfo {author} {\bibfnamefont {D.~V.}\ \bibnamefont
  {Massote}}, \bibinfo {author} {\bibfnamefont {L.~C.}\ \bibnamefont {Campos}},
  \bibinfo {author} {\bibfnamefont {N.~M.}\ \bibnamefont {Barbosa~Neto}},
  \bibinfo {author} {\bibfnamefont {L.~G.}\ \bibnamefont {Can\c{c}ado}},
  \bibinfo {author} {\bibfnamefont {R.~G.}\ \bibnamefont {Lacerda}}, \bibinfo
  {author} {\bibfnamefont {H.}~\bibnamefont {Chacham}}, \bibinfo {author}
  {\bibfnamefont {M.~S.}\ \bibnamefont {Mazzoni}},\ and\ \bibinfo {author}
  {\bibfnamefont {B.~R.}\ \bibnamefont {Neves}},\ }\bibfield  {title} {\bibinfo
  {title} {Room-temperature compression-induced diamondization of few-layer
  graphene},\ }\href@noop {} {\bibfield  {journal} {\bibinfo  {journal} {Adv.
  Mat.}\ }\textbf {\bibinfo {volume} {23}},\ \bibinfo {pages} {3014} (\bibinfo
  {year} {2011})}\BibitemShut {NoStop}%
\bibitem [{\citenamefont {Antipina}\ and\ \citenamefont
  {Sorokin}(2015)}]{antipina2015converting}%
  \BibitemOpen
  \bibfield  {author} {\bibinfo {author} {\bibfnamefont {L.~Y.}\ \bibnamefont
  {Antipina}}\ and\ \bibinfo {author} {\bibfnamefont {P.~B.}\ \bibnamefont
  {Sorokin}},\ }\bibfield  {title} {\bibinfo {title} {Converting chemically
  functionalized few-layer graphene to diamond films: a computational study},\
  }\href@noop {} {\bibfield  {journal} {\bibinfo  {journal} {J. Phys. Chem. C}\
  }\textbf {\bibinfo {volume} {119}},\ \bibinfo {pages} {2828} (\bibinfo {year}
  {2015})}\BibitemShut {NoStop}%
\bibitem [{\citenamefont {Gao}\ \emph {et~al.}(2018)\citenamefont {Gao},
  \citenamefont {Cao}, \citenamefont {Cellini}, \citenamefont {Berger},
  \citenamefont {De~Heer}, \citenamefont {Tosatti}, \citenamefont {Riedo},\
  and\ \citenamefont {Bongiorno}}]{gao2018ultrahard}%
  \BibitemOpen
  \bibfield  {author} {\bibinfo {author} {\bibfnamefont {Y.}~\bibnamefont
  {Gao}}, \bibinfo {author} {\bibfnamefont {T.}~\bibnamefont {Cao}}, \bibinfo
  {author} {\bibfnamefont {F.}~\bibnamefont {Cellini}}, \bibinfo {author}
  {\bibfnamefont {C.}~\bibnamefont {Berger}}, \bibinfo {author} {\bibfnamefont
  {W.~A.}\ \bibnamefont {De~Heer}}, \bibinfo {author} {\bibfnamefont
  {E.}~\bibnamefont {Tosatti}}, \bibinfo {author} {\bibfnamefont
  {E.}~\bibnamefont {Riedo}},\ and\ \bibinfo {author} {\bibfnamefont
  {A.}~\bibnamefont {Bongiorno}},\ }\bibfield  {title} {\bibinfo {title}
  {Ultrahard carbon film from epitaxial two-layer graphene},\ }\href@noop {}
  {\bibfield  {journal} {\bibinfo  {journal} {Nat. Nanotech.}\ }\textbf
  {\bibinfo {volume} {13}},\ \bibinfo {pages} {133} (\bibinfo {year}
  {2018})}\BibitemShut {NoStop}%
\bibitem [{\citenamefont {Piazza}\ \emph
  {et~al.}(2019{\natexlab{a}})\citenamefont {Piazza}, \citenamefont {Gough},
  \citenamefont {Monthioux}, \citenamefont {Puech}, \citenamefont {Gerber},
  \citenamefont {Wiens}, \citenamefont {Paredes},\ and\ \citenamefont
  {Ozoria}}]{piazza2019low}%
  \BibitemOpen
  \bibfield  {author} {\bibinfo {author} {\bibfnamefont {F.}~\bibnamefont
  {Piazza}}, \bibinfo {author} {\bibfnamefont {K.}~\bibnamefont {Gough}},
  \bibinfo {author} {\bibfnamefont {M.}~\bibnamefont {Monthioux}}, \bibinfo
  {author} {\bibfnamefont {P.}~\bibnamefont {Puech}}, \bibinfo {author}
  {\bibfnamefont {I.}~\bibnamefont {Gerber}}, \bibinfo {author} {\bibfnamefont
  {R.}~\bibnamefont {Wiens}}, \bibinfo {author} {\bibfnamefont
  {G.}~\bibnamefont {Paredes}},\ and\ \bibinfo {author} {\bibfnamefont
  {C.}~\bibnamefont {Ozoria}},\ }\bibfield  {title} {\bibinfo {title} {Low
  temperature, pressureless sp2 to sp3 transformation of ultrathin, crystalline
  carbon films},\ }\href@noop {} {\bibfield  {journal} {\bibinfo  {journal}
  {Carbon}\ }\textbf {\bibinfo {volume} {145}},\ \bibinfo {pages} {10}
  (\bibinfo {year} {2019}{\natexlab{a}})}\BibitemShut {NoStop}%
\bibitem [{\citenamefont {Martins}\ \emph {et~al.}(2017)\citenamefont
  {Martins}, \citenamefont {Matos}, \citenamefont {Paschoal}, \citenamefont
  {Freire}, \citenamefont {Andrade}, \citenamefont {Aguiar}, \citenamefont
  {Kong}, \citenamefont {Neves}, \citenamefont {de~Oliveira}, \citenamefont
  {Mazzoni} \emph {et~al.}}]{martins2017raman}%
  \BibitemOpen
  \bibfield  {author} {\bibinfo {author} {\bibfnamefont {L.~G.~P.}\
  \bibnamefont {Martins}}, \bibinfo {author} {\bibfnamefont {M.~J.}\
  \bibnamefont {Matos}}, \bibinfo {author} {\bibfnamefont {A.~R.}\ \bibnamefont
  {Paschoal}}, \bibinfo {author} {\bibfnamefont {P.~T.}\ \bibnamefont
  {Freire}}, \bibinfo {author} {\bibfnamefont {N.~F.}\ \bibnamefont {Andrade}},
  \bibinfo {author} {\bibfnamefont {A.~L.}\ \bibnamefont {Aguiar}}, \bibinfo
  {author} {\bibfnamefont {J.}~\bibnamefont {Kong}}, \bibinfo {author}
  {\bibfnamefont {B.~R.}\ \bibnamefont {Neves}}, \bibinfo {author}
  {\bibfnamefont {A.~B.}\ \bibnamefont {de~Oliveira}}, \bibinfo {author}
  {\bibfnamefont {M.~S.}\ \bibnamefont {Mazzoni}}, \emph {et~al.},\ }\bibfield
  {title} {\bibinfo {title} {Raman evidence for pressure-induced formation of
  diamondene},\ }\href@noop {} {\bibfield  {journal} {\bibinfo  {journal} {Nat.
  Commun.}\ }\textbf {\bibinfo {volume} {8}},\ \bibinfo {pages} {96} (\bibinfo
  {year} {2017})}\BibitemShut {NoStop}%
\bibitem [{\citenamefont {Jorio}\ \emph {et~al.}(2011)\citenamefont {Jorio},
  \citenamefont {Dresselhaus}, \citenamefont {Saito},\ and\ \citenamefont
  {Dresselhaus}}]{jorio2011raman}%
  \BibitemOpen
  \bibfield  {author} {\bibinfo {author} {\bibfnamefont {A.}~\bibnamefont
  {Jorio}}, \bibinfo {author} {\bibfnamefont {M.~S.}\ \bibnamefont
  {Dresselhaus}}, \bibinfo {author} {\bibfnamefont {R.}~\bibnamefont {Saito}},\
  and\ \bibinfo {author} {\bibfnamefont {G.}~\bibnamefont {Dresselhaus}},\
  }\href@noop {} {\emph {\bibinfo {title} {Raman spectroscopy in graphene
  related systems}}}\ (\bibinfo  {publisher} {John Wiley \& Sons},\ \bibinfo
  {year} {2011})\BibitemShut {NoStop}%
\bibitem [{\citenamefont {Huang}\ \emph {et~al.}(2009)\citenamefont {Huang},
  \citenamefont {Yan}, \citenamefont {Chen}, \citenamefont {Song},
  \citenamefont {Heinz},\ and\ \citenamefont {Hone}}]{huang2009phonon}%
  \BibitemOpen
  \bibfield  {author} {\bibinfo {author} {\bibfnamefont {M.}~\bibnamefont
  {Huang}}, \bibinfo {author} {\bibfnamefont {H.}~\bibnamefont {Yan}}, \bibinfo
  {author} {\bibfnamefont {C.}~\bibnamefont {Chen}}, \bibinfo {author}
  {\bibfnamefont {D.}~\bibnamefont {Song}}, \bibinfo {author} {\bibfnamefont
  {T.~F.}\ \bibnamefont {Heinz}},\ and\ \bibinfo {author} {\bibfnamefont
  {J.}~\bibnamefont {Hone}},\ }\bibfield  {title} {\bibinfo {title} {Phonon
  softening and crystallographic orientation of strained graphene studied by
  raman spectroscopy},\ }\href@noop {} {\bibfield  {journal} {\bibinfo
  {journal} {PNAS}\ }\textbf {\bibinfo {volume} {106}},\ \bibinfo {pages}
  {7304} (\bibinfo {year} {2009})}\BibitemShut {NoStop}%
\bibitem [{\citenamefont {Proctor}\ \emph {et~al.}(2009)\citenamefont
  {Proctor}, \citenamefont {Gregoryanz}, \citenamefont {Novoselov},
  \citenamefont {Lotya}, \citenamefont {Coleman},\ and\ \citenamefont
  {Halsall}}]{proctor2009high}%
  \BibitemOpen
  \bibfield  {author} {\bibinfo {author} {\bibfnamefont {J.~E.}\ \bibnamefont
  {Proctor}}, \bibinfo {author} {\bibfnamefont {E.}~\bibnamefont {Gregoryanz}},
  \bibinfo {author} {\bibfnamefont {K.~S.}\ \bibnamefont {Novoselov}}, \bibinfo
  {author} {\bibfnamefont {M.}~\bibnamefont {Lotya}}, \bibinfo {author}
  {\bibfnamefont {J.~N.}\ \bibnamefont {Coleman}},\ and\ \bibinfo {author}
  {\bibfnamefont {M.~P.}\ \bibnamefont {Halsall}},\ }\bibfield  {title}
  {\bibinfo {title} {High-pressure raman spectroscopy of graphene},\
  }\href@noop {} {\bibfield  {journal} {\bibinfo  {journal} {Phys. Rev. B}\
  }\textbf {\bibinfo {volume} {80}},\ \bibinfo {pages} {073408} (\bibinfo
  {year} {2009})}\BibitemShut {NoStop}%
\bibitem [{\citenamefont {Filintoglou}\ \emph {et~al.}(2013)\citenamefont
  {Filintoglou}, \citenamefont {Papadopoulos}, \citenamefont {Arvanitidis},
  \citenamefont {Christofilos}, \citenamefont {Frank}, \citenamefont {Kalbac},
  \citenamefont {Parthenios}, \citenamefont {Kalosakas}, \citenamefont
  {Galiotis},\ and\ \citenamefont {Papagelis}}]{filintoglou2013raman}%
  \BibitemOpen
  \bibfield  {author} {\bibinfo {author} {\bibfnamefont {K.}~\bibnamefont
  {Filintoglou}}, \bibinfo {author} {\bibfnamefont {N.}~\bibnamefont
  {Papadopoulos}}, \bibinfo {author} {\bibfnamefont {J.}~\bibnamefont
  {Arvanitidis}}, \bibinfo {author} {\bibfnamefont {D.}~\bibnamefont
  {Christofilos}}, \bibinfo {author} {\bibfnamefont {O.}~\bibnamefont {Frank}},
  \bibinfo {author} {\bibfnamefont {M.}~\bibnamefont {Kalbac}}, \bibinfo
  {author} {\bibfnamefont {J.}~\bibnamefont {Parthenios}}, \bibinfo {author}
  {\bibfnamefont {G.}~\bibnamefont {Kalosakas}}, \bibinfo {author}
  {\bibfnamefont {C.}~\bibnamefont {Galiotis}},\ and\ \bibinfo {author}
  {\bibfnamefont {K.}~\bibnamefont {Papagelis}},\ }\bibfield  {title} {\bibinfo
  {title} {Raman spectroscopy of graphene at high pressure: Effects of the
  substrate and the pressure transmitting media},\ }\href@noop {} {\bibfield
  {journal} {\bibinfo  {journal} {Phys. Rev. B}\ }\textbf {\bibinfo {volume}
  {88}},\ \bibinfo {pages} {045418} (\bibinfo {year} {2013})}\BibitemShut
  {NoStop}%
\bibitem [{\citenamefont {Hanfland}\ \emph {et~al.}(1989)\citenamefont
  {Hanfland}, \citenamefont {Beister},\ and\ \citenamefont
  {Syassen}}]{hanfland1989graphite}%
  \BibitemOpen
  \bibfield  {author} {\bibinfo {author} {\bibfnamefont {M.}~\bibnamefont
  {Hanfland}}, \bibinfo {author} {\bibfnamefont {H.}~\bibnamefont {Beister}},\
  and\ \bibinfo {author} {\bibfnamefont {K.}~\bibnamefont {Syassen}},\
  }\bibfield  {title} {\bibinfo {title} {Graphite under pressure: Equation of
  state and first-order raman modes},\ }\href@noop {} {\bibfield  {journal}
  {\bibinfo  {journal} {Phys. Rev. B}\ }\textbf {\bibinfo {volume} {39}},\
  \bibinfo {pages} {12598} (\bibinfo {year} {1989})}\BibitemShut {NoStop}%
\bibitem [{\citenamefont {Amsler}\ \emph {et~al.}(2012)\citenamefont {Amsler},
  \citenamefont {Flores-Livas}, \citenamefont {Lehtovaara}, \citenamefont
  {Balima}, \citenamefont {Ghasemi}, \citenamefont {Machon}, \citenamefont
  {Pailh{\`e}s}, \citenamefont {Willand}, \citenamefont {Caliste},
  \citenamefont {Botti} \emph {et~al.}}]{amsler2012crystal}%
  \BibitemOpen
  \bibfield  {author} {\bibinfo {author} {\bibfnamefont {M.}~\bibnamefont
  {Amsler}}, \bibinfo {author} {\bibfnamefont {J.~A.}\ \bibnamefont
  {Flores-Livas}}, \bibinfo {author} {\bibfnamefont {L.}~\bibnamefont
  {Lehtovaara}}, \bibinfo {author} {\bibfnamefont {F.}~\bibnamefont {Balima}},
  \bibinfo {author} {\bibfnamefont {S.~A.}\ \bibnamefont {Ghasemi}}, \bibinfo
  {author} {\bibfnamefont {D.}~\bibnamefont {Machon}}, \bibinfo {author}
  {\bibfnamefont {S.}~\bibnamefont {Pailh{\`e}s}}, \bibinfo {author}
  {\bibfnamefont {A.}~\bibnamefont {Willand}}, \bibinfo {author} {\bibfnamefont
  {D.}~\bibnamefont {Caliste}}, \bibinfo {author} {\bibfnamefont
  {S.}~\bibnamefont {Botti}}, \emph {et~al.},\ }\bibfield  {title} {\bibinfo
  {title} {Crystal structure of cold compressed graphite},\ }\href@noop {}
  {\bibfield  {journal} {\bibinfo  {journal} {Phys. Rev. Letters}\ }\textbf
  {\bibinfo {volume} {108}},\ \bibinfo {pages} {065501} (\bibinfo {year}
  {2012})}\BibitemShut {NoStop}%
\bibitem [{\citenamefont {Wang}\ \emph {et~al.}(2012)\citenamefont {Wang},
  \citenamefont {Panzik}, \citenamefont {Kiefer},\ and\ \citenamefont
  {Lee}}]{wang2012crystal}%
  \BibitemOpen
  \bibfield  {author} {\bibinfo {author} {\bibfnamefont {Y.}~\bibnamefont
  {Wang}}, \bibinfo {author} {\bibfnamefont {J.~E.}\ \bibnamefont {Panzik}},
  \bibinfo {author} {\bibfnamefont {B.}~\bibnamefont {Kiefer}},\ and\ \bibinfo
  {author} {\bibfnamefont {K.~K.}\ \bibnamefont {Lee}},\ }\bibfield  {title}
  {\bibinfo {title} {Crystal structure of graphite under room-temperature
  compression and decompression},\ }\href@noop {} {\bibfield  {journal}
  {\bibinfo  {journal} {Sci. Rep.}\ }\textbf {\bibinfo {volume} {2}},\ \bibinfo
  {pages} {520} (\bibinfo {year} {2012})}\BibitemShut {NoStop}%
\bibitem [{\citenamefont {Ferreira}\ \emph {et~al.}(2010)\citenamefont
  {Ferreira}, \citenamefont {Moutinho}, \citenamefont {Stavale}, \citenamefont
  {Lucchese}, \citenamefont {Capaz}, \citenamefont {Achete},\ and\
  \citenamefont {Jorio}}]{ferreira2010evolution}%
  \BibitemOpen
  \bibfield  {author} {\bibinfo {author} {\bibfnamefont {E.~M.}\ \bibnamefont
  {Ferreira}}, \bibinfo {author} {\bibfnamefont {M.~V.}\ \bibnamefont
  {Moutinho}}, \bibinfo {author} {\bibfnamefont {F.}~\bibnamefont {Stavale}},
  \bibinfo {author} {\bibfnamefont {M.}~\bibnamefont {Lucchese}}, \bibinfo
  {author} {\bibfnamefont {R.~B.}\ \bibnamefont {Capaz}}, \bibinfo {author}
  {\bibfnamefont {C.}~\bibnamefont {Achete}},\ and\ \bibinfo {author}
  {\bibfnamefont {A.}~\bibnamefont {Jorio}},\ }\bibfield  {title} {\bibinfo
  {title} {Evolution of the raman spectra from single-, few-, and many-layer
  graphene with increasing disorder},\ }\href@noop {} {\bibfield  {journal}
  {\bibinfo  {journal} {Phys. Rev. B}\ }\textbf {\bibinfo {volume} {82}},\
  \bibinfo {pages} {125429} (\bibinfo {year} {2010})}\BibitemShut {NoStop}%
\bibitem [{\citenamefont {Elias}\ \emph {et~al.}(2009)\citenamefont {Elias},
  \citenamefont {Nair}, \citenamefont {Mohiuddin}, \citenamefont {Morozov},
  \citenamefont {Blake}, \citenamefont {Halsall}, \citenamefont {Ferrari},
  \citenamefont {Boukhvalov}, \citenamefont {Katsnelson}, \citenamefont {Geim}
  \emph {et~al.}}]{elias2009control}%
  \BibitemOpen
  \bibfield  {author} {\bibinfo {author} {\bibfnamefont {D.~C.}\ \bibnamefont
  {Elias}}, \bibinfo {author} {\bibfnamefont {R.~R.}\ \bibnamefont {Nair}},
  \bibinfo {author} {\bibfnamefont {T.}~\bibnamefont {Mohiuddin}}, \bibinfo
  {author} {\bibfnamefont {S.}~\bibnamefont {Morozov}}, \bibinfo {author}
  {\bibfnamefont {P.}~\bibnamefont {Blake}}, \bibinfo {author} {\bibfnamefont
  {M.}~\bibnamefont {Halsall}}, \bibinfo {author} {\bibfnamefont {A.~C.}\
  \bibnamefont {Ferrari}}, \bibinfo {author} {\bibfnamefont {D.}~\bibnamefont
  {Boukhvalov}}, \bibinfo {author} {\bibfnamefont {M.}~\bibnamefont
  {Katsnelson}}, \bibinfo {author} {\bibfnamefont {A.}~\bibnamefont {Geim}},
  \emph {et~al.},\ }\bibfield  {title} {\bibinfo {title} {Control of graphene's
  properties by reversible hydrogenation: evidence for graphane},\ }\href@noop
  {} {\bibfield  {journal} {\bibinfo  {journal} {Science}\ }\textbf {\bibinfo
  {volume} {323}},\ \bibinfo {pages} {610} (\bibinfo {year}
  {2009})}\BibitemShut {NoStop}%
\bibitem [{\citenamefont {Yagi}\ \emph {et~al.}(1992)\citenamefont {Yagi},
  \citenamefont {Utsumi}, \citenamefont {Yamakata}, \citenamefont {Kikegawa},\
  and\ \citenamefont {Shimomura}}]{yagi1992high}%
  \BibitemOpen
  \bibfield  {author} {\bibinfo {author} {\bibfnamefont {T.}~\bibnamefont
  {Yagi}}, \bibinfo {author} {\bibfnamefont {W.}~\bibnamefont {Utsumi}},
  \bibinfo {author} {\bibfnamefont {M.-a.}\ \bibnamefont {Yamakata}}, \bibinfo
  {author} {\bibfnamefont {T.}~\bibnamefont {Kikegawa}},\ and\ \bibinfo
  {author} {\bibfnamefont {O.}~\bibnamefont {Shimomura}},\ }\bibfield  {title}
  {\bibinfo {title} {High-pressure in situ x-ray-diffraction study of the phase
  transformation from graphite to hexagonal diamond at room temperature},\
  }\href@noop {} {\bibfield  {journal} {\bibinfo  {journal} {Phys. Rev. B}\
  }\textbf {\bibinfo {volume} {46}},\ \bibinfo {pages} {6031} (\bibinfo {year}
  {1992})}\BibitemShut {NoStop}%
\bibitem [{\citenamefont {Paul}\ and\ \citenamefont
  {Momeni}(2019)}]{paul2019mechanochemistry}%
  \BibitemOpen
  \bibfield  {author} {\bibinfo {author} {\bibfnamefont {S.}~\bibnamefont
  {Paul}}\ and\ \bibinfo {author} {\bibfnamefont {K.}~\bibnamefont {Momeni}},\
  }\bibfield  {title} {\bibinfo {title} {Mechanochemistry of stable diamane and
  atomically thin diamond films synthesis from bi-and multilayer graphene: A
  computational study},\ }\href@noop {} {\bibfield  {journal} {\bibinfo
  {journal} {J. Phys. Chem. C}\ } (\bibinfo {year} {2019})}\BibitemShut
  {NoStop}%
\bibitem [{\citenamefont {Piazza}\ \emph
  {et~al.}(2019{\natexlab{b}})\citenamefont {Piazza}, \citenamefont
  {Monthioux}, \citenamefont {Puech},\ and\ \citenamefont
  {Gerber}}]{piazza2019towards}%
  \BibitemOpen
  \bibfield  {author} {\bibinfo {author} {\bibfnamefont {F.}~\bibnamefont
  {Piazza}}, \bibinfo {author} {\bibfnamefont {M.}~\bibnamefont {Monthioux}},
  \bibinfo {author} {\bibfnamefont {P.}~\bibnamefont {Puech}},\ and\ \bibinfo
  {author} {\bibfnamefont {I.~C.}\ \bibnamefont {Gerber}},\ }\bibfield  {title}
  {\bibinfo {title} {Towards a better understanding of the structure of
  diamano\"{i}ds and diamano\"{i}d/graphene hybrids},\ }\href@noop {}
  {\bibfield  {journal} {\bibinfo  {journal} {arXiv preprint arXiv:1907.09033}\
  } (\bibinfo {year} {2019}{\natexlab{b}})}\BibitemShut {NoStop}%
\bibitem [{\citenamefont {Clark}\ \emph {et~al.}(2013)\citenamefont {Clark},
  \citenamefont {Jeon}, \citenamefont {Chen},\ and\ \citenamefont
  {Yoo}}]{clark2013few}%
  \BibitemOpen
  \bibfield  {author} {\bibinfo {author} {\bibfnamefont {S.}~\bibnamefont
  {Clark}}, \bibinfo {author} {\bibfnamefont {K.-J.}\ \bibnamefont {Jeon}},
  \bibinfo {author} {\bibfnamefont {J.-Y.}\ \bibnamefont {Chen}},\ and\
  \bibinfo {author} {\bibfnamefont {C.-S.}\ \bibnamefont {Yoo}},\ }\bibfield
  {title} {\bibinfo {title} {Few-layer graphene under high pressure: Raman and
  x-ray diffraction studies},\ }\href@noop {} {\bibfield  {journal} {\bibinfo
  {journal} {Solid State Commun.}\ }\textbf {\bibinfo {volume} {154}},\
  \bibinfo {pages} {15} (\bibinfo {year} {2013})}\BibitemShut {NoStop}%
\bibitem [{\citenamefont {Cellini}\ \emph {et~al.}(2019)\citenamefont
  {Cellini}, \citenamefont {Lavini}, \citenamefont {Berger}, \citenamefont
  {de~Heer},\ and\ \citenamefont {Riedo}}]{cellini2019layer}%
  \BibitemOpen
  \bibfield  {author} {\bibinfo {author} {\bibfnamefont {F.}~\bibnamefont
  {Cellini}}, \bibinfo {author} {\bibfnamefont {F.}~\bibnamefont {Lavini}},
  \bibinfo {author} {\bibfnamefont {C.}~\bibnamefont {Berger}}, \bibinfo
  {author} {\bibfnamefont {W.}~\bibnamefont {de~Heer}},\ and\ \bibinfo {author}
  {\bibfnamefont {E.}~\bibnamefont {Riedo}},\ }\bibfield  {title} {\bibinfo
  {title} {Layer dependence of graphene-diamene phase transition in epitaxial
  and exfoliated few-layer graphene using machine learning},\ }\href@noop {}
  {\bibfield  {journal} {\bibinfo  {journal} {2D Materials}\ } (\bibinfo {year}
  {2019})}\BibitemShut {NoStop}%
\bibitem [{\citenamefont {Ke}\ \emph {et~al.}(2019)\citenamefont {Ke},
  \citenamefont {Chen}, \citenamefont {Yin}, \citenamefont {Yan}, \citenamefont
  {Zhang}, \citenamefont {Liu}, \citenamefont {John}, \citenamefont {Wu},
  \citenamefont {Mao},\ and\ \citenamefont {Chen}}]{ke2019large}%
  \BibitemOpen
  \bibfield  {author} {\bibinfo {author} {\bibfnamefont {F.}~\bibnamefont
  {Ke}}, \bibinfo {author} {\bibfnamefont {Y.}~\bibnamefont {Chen}}, \bibinfo
  {author} {\bibfnamefont {K.}~\bibnamefont {Yin}}, \bibinfo {author}
  {\bibfnamefont {J.}~\bibnamefont {Yan}}, \bibinfo {author} {\bibfnamefont
  {H.}~\bibnamefont {Zhang}}, \bibinfo {author} {\bibfnamefont
  {Z.}~\bibnamefont {Liu}}, \bibinfo {author} {\bibfnamefont {S.~T.}\
  \bibnamefont {John}}, \bibinfo {author} {\bibfnamefont {J.}~\bibnamefont
  {Wu}}, \bibinfo {author} {\bibfnamefont {H.-k.}\ \bibnamefont {Mao}},\ and\
  \bibinfo {author} {\bibfnamefont {B.}~\bibnamefont {Chen}},\ }\bibfield
  {title} {\bibinfo {title} {Large bandgap of pressurized trilayer graphene},\
  }\href@noop {} {\bibfield  {journal} {\bibinfo  {journal} {PNAS}\ }\textbf
  {\bibinfo {volume} {116}},\ \bibinfo {pages} {9186} (\bibinfo {year}
  {2019})}\BibitemShut {NoStop}%
\bibitem [{\citenamefont {Srivastava}\ and\ \citenamefont
  {Fahad}(2019)}]{srivastava2019synthesis}%
  \BibitemOpen
  \bibfield  {author} {\bibinfo {author} {\bibfnamefont {A.}~\bibnamefont
  {Srivastava}}\ and\ \bibinfo {author} {\bibfnamefont {M.~S.}\ \bibnamefont
  {Fahad}},\ }\bibfield  {title} {\bibinfo {title} {Synthesis of
  two-dimensional diamond from graphene on copper},\ }in\ \href@noop {} {\emph
  {\bibinfo {booktitle} {2019 Electron Devices Technology and Manufacturing
  Conference (EDTM)}}}\ (\bibinfo {organization} {IEEE},\ \bibinfo {year}
  {2019})\ pp.\ \bibinfo {pages} {273--275}\BibitemShut {NoStop}%
\bibitem [{\citenamefont {Bakharev}\ \emph {et~al.}(2019)\citenamefont
  {Bakharev}, \citenamefont {Huang}, \citenamefont {Saxena}, \citenamefont
  {Lee}, \citenamefont {Joo}, \citenamefont {Park}, \citenamefont {Dong},
  \citenamefont {Camacho-Mojica}, \citenamefont {Jin}, \citenamefont {Kwon}
  \emph {et~al.}}]{bakharev2019chemically}%
  \BibitemOpen
  \bibfield  {author} {\bibinfo {author} {\bibfnamefont {P.~V.}\ \bibnamefont
  {Bakharev}}, \bibinfo {author} {\bibfnamefont {M.}~\bibnamefont {Huang}},
  \bibinfo {author} {\bibfnamefont {M.}~\bibnamefont {Saxena}}, \bibinfo
  {author} {\bibfnamefont {S.~W.}\ \bibnamefont {Lee}}, \bibinfo {author}
  {\bibfnamefont {S.~H.}\ \bibnamefont {Joo}}, \bibinfo {author} {\bibfnamefont
  {S.~O.}\ \bibnamefont {Park}}, \bibinfo {author} {\bibfnamefont
  {J.}~\bibnamefont {Dong}}, \bibinfo {author} {\bibfnamefont {D.}~\bibnamefont
  {Camacho-Mojica}}, \bibinfo {author} {\bibfnamefont {S.}~\bibnamefont {Jin}},
  \bibinfo {author} {\bibfnamefont {Y.}~\bibnamefont {Kwon}}, \emph {et~al.},\
  }\bibfield  {title} {\bibinfo {title} {Chemically induced transformation of
  cvd-grown bilayer graphene into single layer diamond},\ }\href@noop {}
  {\bibfield  {journal} {\bibinfo  {journal} {arXiv preprint arXiv:1901.02131}\
  } (\bibinfo {year} {2019})}\BibitemShut {NoStop}%
\bibitem [{\citenamefont {Wang}\ \emph {et~al.}(2015)\citenamefont {Wang},
  \citenamefont {Yang}, \citenamefont {Chen}, \citenamefont {Watanabe},
  \citenamefont {Taniguchi}, \citenamefont {Churchill},\ and\ \citenamefont
  {Jarillo-Herrero}}]{wang2015electronic}%
  \BibitemOpen
  \bibfield  {author} {\bibinfo {author} {\bibfnamefont {J.~I.-J.}\
  \bibnamefont {Wang}}, \bibinfo {author} {\bibfnamefont {Y.}~\bibnamefont
  {Yang}}, \bibinfo {author} {\bibfnamefont {Y.-A.}\ \bibnamefont {Chen}},
  \bibinfo {author} {\bibfnamefont {K.}~\bibnamefont {Watanabe}}, \bibinfo
  {author} {\bibfnamefont {T.}~\bibnamefont {Taniguchi}}, \bibinfo {author}
  {\bibfnamefont {H.~O.}\ \bibnamefont {Churchill}},\ and\ \bibinfo {author}
  {\bibfnamefont {P.}~\bibnamefont {Jarillo-Herrero}},\ }\bibfield  {title}
  {\bibinfo {title} {Electronic transport of encapsulated graphene and wse2
  devices fabricated by pick-up of prepatterned hbn},\ }\href@noop {}
  {\bibfield  {journal} {\bibinfo  {journal} {Nano Lett.}\ }\textbf {\bibinfo
  {volume} {15}},\ \bibinfo {pages} {1898} (\bibinfo {year}
  {2015})}\BibitemShut {NoStop}%
\bibitem [{\citenamefont {Fernandes}\ \emph {et~al.}(2019)\citenamefont
  {Fernandes}, \citenamefont {Miquita}, \citenamefont {Can\c{c}ado},\ and\
  \citenamefont {Neves}}]{Thales2019}%
  \BibitemOpen
  \bibfield  {author} {\bibinfo {author} {\bibfnamefont {T.~F.~D.}\
  \bibnamefont {Fernandes}}, \bibinfo {author} {\bibfnamefont {D.~R.}\
  \bibnamefont {Miquita}}, \bibinfo {author} {\bibfnamefont {L.~G.}\
  \bibnamefont {Can\c{c}ado}},\ and\ \bibinfo {author} {\bibfnamefont
  {B.~R.~A.}\ \bibnamefont {Neves}},\ }\href@noop {} {\bibfield  {journal}
  {\bibinfo  {journal} {In preparation}\ } (\bibinfo {year}
  {2019})}\BibitemShut {NoStop}%
\bibitem [{\citenamefont {Mao}\ \emph {et~al.}(1986)\citenamefont {Mao},
  \citenamefont {Xu},\ and\ \citenamefont {Bell}}]{mao1986calibration}%
  \BibitemOpen
  \bibfield  {author} {\bibinfo {author} {\bibfnamefont {H.}~\bibnamefont
  {Mao}}, \bibinfo {author} {\bibfnamefont {J.-A.}\ \bibnamefont {Xu}},\ and\
  \bibinfo {author} {\bibfnamefont {P.}~\bibnamefont {Bell}},\ }\bibfield
  {title} {\bibinfo {title} {Calibration of the ruby pressure gauge to 800 kbar
  under quasi-hydrostatic conditions},\ }\href@noop {} {\bibfield  {journal}
  {\bibinfo  {journal} {J. Geophys. Res.: Sol. Earth}\ }\textbf {\bibinfo
  {volume} {91}},\ \bibinfo {pages} {4673} (\bibinfo {year}
  {1986})}\BibitemShut {NoStop}%
\bibitem [{\citenamefont {Silva}\ \emph {et~al.}(2019)\citenamefont {Silva},
  \citenamefont {Campos}, \citenamefont {Fernandes}, \citenamefont {Nunes},
  \citenamefont {Miranda}, \citenamefont {Rabelo}, \citenamefont {Vilela~Neto},
  \citenamefont {Jorio},\ and\ \citenamefont {Can\c{c}ado}}]{Diego2019}%
  \BibitemOpen
  \bibfield  {author} {\bibinfo {author} {\bibfnamefont {D.~L.}\ \bibnamefont
  {Silva}}, \bibinfo {author} {\bibfnamefont {J.~L.~E.}\ \bibnamefont
  {Campos}}, \bibinfo {author} {\bibfnamefont {T.~F.~D.}\ \bibnamefont
  {Fernandes}}, \bibinfo {author} {\bibfnamefont {J.}~\bibnamefont {Nunes}},
  \bibinfo {author} {\bibfnamefont {H.}~\bibnamefont {Miranda}}, \bibinfo
  {author} {\bibfnamefont {C.}~\bibnamefont {Rabelo}}, \bibinfo {author}
  {\bibfnamefont {O.~P.}\ \bibnamefont {Vilela~Neto}}, \bibinfo {author}
  {\bibfnamefont {A.}~\bibnamefont {Jorio}},\ and\ \bibinfo {author}
  {\bibfnamefont {L.~G.}\ \bibnamefont {Can\c{c}ado}},\ }\bibfield  {title}
  {\bibinfo {title} {Determination of statistical distributions of number of
  layers in graphene systems by raman spectroscopy},\ }\href@noop {} {\bibfield
   {journal} {\bibinfo  {journal} {In preparation}\ } (\bibinfo {year}
  {2019})}\BibitemShut {NoStop}%
\bibitem [{\citenamefont {Plimpton}(1995)}]{plimpton1995fast}%
  \BibitemOpen
  \bibfield  {author} {\bibinfo {author} {\bibfnamefont {S.}~\bibnamefont
  {Plimpton}},\ }\bibfield  {title} {\bibinfo {title} {Fast parallel algorithms
  for short-range molecular dynamics},\ }\href@noop {} {\bibfield  {journal}
  {\bibinfo  {journal} {J. Comp. Phys.}\ }\textbf {\bibinfo {volume} {117}},\
  \bibinfo {pages} {1} (\bibinfo {year} {1995})}\BibitemShut {NoStop}%
\bibitem [{\citenamefont {Stuart}\ \emph {et~al.}(2000)\citenamefont {Stuart},
  \citenamefont {Tutein},\ and\ \citenamefont {Harrison}}]{stuart2000reactive}%
  \BibitemOpen
  \bibfield  {author} {\bibinfo {author} {\bibfnamefont {S.~J.}\ \bibnamefont
  {Stuart}}, \bibinfo {author} {\bibfnamefont {A.~B.}\ \bibnamefont {Tutein}},\
  and\ \bibinfo {author} {\bibfnamefont {J.~A.}\ \bibnamefont {Harrison}},\
  }\bibfield  {title} {\bibinfo {title} {A reactive potential for hydrocarbons
  with intermolecular interactions},\ }\href@noop {} {\bibfield  {journal}
  {\bibinfo  {journal} {J. Chem. Phys.}\ }\textbf {\bibinfo {volume} {112}},\
  \bibinfo {pages} {6472} (\bibinfo {year} {2000})}\BibitemShut {NoStop}%
\bibitem [{\citenamefont {Nos{\'e}}(1984)}]{nose1984unified}%
  \BibitemOpen
  \bibfield  {author} {\bibinfo {author} {\bibfnamefont {S.}~\bibnamefont
  {Nos{\'e}}},\ }\bibfield  {title} {\bibinfo {title} {A unified formulation of
  the constant temperature molecular dynamics methods},\ }\href@noop {}
  {\bibfield  {journal} {\bibinfo  {journal} {J. Chem. Phys.}\ }\textbf
  {\bibinfo {volume} {81}},\ \bibinfo {pages} {511} (\bibinfo {year}
  {1984})}\BibitemShut {NoStop}%
\bibitem [{\citenamefont {Hohenberg}\ and\ \citenamefont
  {Kohn}(1964)}]{hohenberg1964inhomogeneous}%
  \BibitemOpen
  \bibfield  {author} {\bibinfo {author} {\bibfnamefont {P.}~\bibnamefont
  {Hohenberg}}\ and\ \bibinfo {author} {\bibfnamefont {W.}~\bibnamefont
  {Kohn}},\ }\bibfield  {title} {\bibinfo {title} {Inhomogeneous electron
  gas},\ }\href@noop {} {\bibfield  {journal} {\bibinfo  {journal} {Phys.
  Rev.}\ }\textbf {\bibinfo {volume} {136}},\ \bibinfo {pages} {B864} (\bibinfo
  {year} {1964})}\BibitemShut {NoStop}%
\bibitem [{\citenamefont {Kohn}\ and\ \citenamefont
  {Sham}(1965)}]{kohn1965self}%
  \BibitemOpen
  \bibfield  {author} {\bibinfo {author} {\bibfnamefont {W.}~\bibnamefont
  {Kohn}}\ and\ \bibinfo {author} {\bibfnamefont {L.~J.}\ \bibnamefont
  {Sham}},\ }\bibfield  {title} {\bibinfo {title} {Self-consistent equations
  including exchange and correlation effects},\ }\href@noop {} {\bibfield
  {journal} {\bibinfo  {journal} {Phys. Rev.}\ }\textbf {\bibinfo {volume}
  {140}},\ \bibinfo {pages} {A1133} (\bibinfo {year} {1965})}\BibitemShut
  {NoStop}%
\bibitem [{\citenamefont {Soler}\ \emph {et~al.}(2002)\citenamefont {Soler},
  \citenamefont {Artacho}, \citenamefont {Gale}, \citenamefont {Garc{\'\i}a},
  \citenamefont {Junquera}, \citenamefont {Ordej{\'o}n},\ and\ \citenamefont
  {S{\'a}nchez-Portal}}]{soler2002siesta}%
  \BibitemOpen
  \bibfield  {author} {\bibinfo {author} {\bibfnamefont {J.~M.}\ \bibnamefont
  {Soler}}, \bibinfo {author} {\bibfnamefont {E.}~\bibnamefont {Artacho}},
  \bibinfo {author} {\bibfnamefont {J.~D.}\ \bibnamefont {Gale}}, \bibinfo
  {author} {\bibfnamefont {A.}~\bibnamefont {Garc{\'\i}a}}, \bibinfo {author}
  {\bibfnamefont {J.}~\bibnamefont {Junquera}}, \bibinfo {author}
  {\bibfnamefont {P.}~\bibnamefont {Ordej{\'o}n}},\ and\ \bibinfo {author}
  {\bibfnamefont {D.}~\bibnamefont {S{\'a}nchez-Portal}},\ }\bibfield  {title}
  {\bibinfo {title} {The siesta method for ab initio order-n materials
  simulation},\ }\href@noop {} {\bibfield  {journal} {\bibinfo  {journal} {J.
  Phys. Condens. Matter}\ }\textbf {\bibinfo {volume} {14}},\ \bibinfo {pages}
  {2745} (\bibinfo {year} {2002})}\BibitemShut {NoStop}%
\bibitem [{\citenamefont {Troullier}\ and\ \citenamefont
  {Martins}(1991)}]{troullier1991efficient}%
  \BibitemOpen
  \bibfield  {author} {\bibinfo {author} {\bibfnamefont {N.}~\bibnamefont
  {Troullier}}\ and\ \bibinfo {author} {\bibfnamefont {J.~L.}\ \bibnamefont
  {Martins}},\ }\bibfield  {title} {\bibinfo {title} {Efficient
  pseudopotentials for plane-wave calculations},\ }\href@noop {} {\bibfield
  {journal} {\bibinfo  {journal} {Phys. Rev. B}\ }\textbf {\bibinfo {volume}
  {43}},\ \bibinfo {pages} {1993} (\bibinfo {year} {1991})}\BibitemShut
  {NoStop}%
\bibitem [{\citenamefont {Kleinman}\ and\ \citenamefont
  {Bylander}(1982)}]{kleinman1982efficacious}%
  \BibitemOpen
  \bibfield  {author} {\bibinfo {author} {\bibfnamefont {L.}~\bibnamefont
  {Kleinman}}\ and\ \bibinfo {author} {\bibfnamefont {D.}~\bibnamefont
  {Bylander}},\ }\bibfield  {title} {\bibinfo {title} {Efficacious form for
  model pseudopotentials},\ }\href@noop {} {\bibfield  {journal} {\bibinfo
  {journal} {Phys. Rev. Lett.}\ }\textbf {\bibinfo {volume} {48}},\ \bibinfo
  {pages} {1425} (\bibinfo {year} {1982})}\BibitemShut {NoStop}%
\bibitem [{\citenamefont {Perdew}\ \emph {et~al.}(1996)\citenamefont {Perdew},
  \citenamefont {Burke},\ and\ \citenamefont
  {Ernzerhof}}]{perdew1996generalized}%
  \BibitemOpen
  \bibfield  {author} {\bibinfo {author} {\bibfnamefont {J.~P.}\ \bibnamefont
  {Perdew}}, \bibinfo {author} {\bibfnamefont {K.}~\bibnamefont {Burke}},\ and\
  \bibinfo {author} {\bibfnamefont {M.}~\bibnamefont {Ernzerhof}},\ }\bibfield
  {title} {\bibinfo {title} {Generalized gradient approximation made simple},\
  }\href@noop {} {\bibfield  {journal} {\bibinfo  {journal} {Phys. Rev. Lett.}\
  }\textbf {\bibinfo {volume} {77}},\ \bibinfo {pages} {3865} (\bibinfo {year}
  {1996})}\BibitemShut {NoStop}%
\bibitem [{\citenamefont {Kokalj}(2003)}]{kokalj2003computer}%
  \BibitemOpen
  \bibfield  {author} {\bibinfo {author} {\bibfnamefont {A.}~\bibnamefont
  {Kokalj}},\ }\bibfield  {title} {\bibinfo {title} {Computer graphics and
  graphical user interfaces as tools in simulations of matter at the atomic
  scale},\ }\href@noop {} {\bibfield  {journal} {\bibinfo  {journal} {Comput.
  Mater. Science}\ }\textbf {\bibinfo {volume} {28}},\ \bibinfo {pages} {155}
  (\bibinfo {year} {2003})}\BibitemShut {NoStop}%
\bibitem [{\citenamefont {Izumi}\ and\ \citenamefont
  {Momma}(2007)}]{izumi2007}%
  \BibitemOpen
  \bibfield  {author} {\bibinfo {author} {\bibfnamefont {F.}~\bibnamefont
  {Izumi}}\ and\ \bibinfo {author} {\bibfnamefont {K.}~\bibnamefont {Momma}},\
  }\bibfield  {title} {\bibinfo {title} {Three-dimensional visualization in
  powder diffraction},\ }in\ \href
  {https://doi.org/10.4028/www.scientific.net/SSP.130.15} {\emph {\bibinfo
  {booktitle} {Applied Crystallography XX}}},\ \bibinfo {series} {Solid State
  Phenomena}, Vol.\ \bibinfo {volume} {130}\ (\bibinfo  {publisher} {Trans Tech
  Publications Ltd},\ \bibinfo {year} {2007})\ pp.\ \bibinfo {pages}
  {15--20}\BibitemShut {NoStop}%
\bibitem [{\citenamefont {Momma}\ and\ \citenamefont
  {Izumi}(2011)}]{Momma:db5098}%
  \BibitemOpen
  \bibfield  {author} {\bibinfo {author} {\bibfnamefont {K.}~\bibnamefont
  {Momma}}\ and\ \bibinfo {author} {\bibfnamefont {F.}~\bibnamefont {Izumi}},\
  }\bibfield  {title} {\bibinfo {title} {{{\it VESTA3} for three-dimensional
  visualization of crystal, volumetric and morphology data}},\ }\href
  {https://doi.org/10.1107/S0021889811038970} {\bibfield  {journal} {\bibinfo
  {journal} {J. Appl. Crystallogr.}\ }\textbf {\bibinfo {volume} {44}},\
  \bibinfo {pages} {1272} (\bibinfo {year} {2011})}\BibitemShut {NoStop}%
\bibitem [{\citenamefont {Piermarini}\ \emph {et~al.}(1973)\citenamefont
  {Piermarini}, \citenamefont {Block},\ and\ \citenamefont
  {Barnett}}]{piermarini1973hydrostatic}%
  \BibitemOpen
  \bibfield  {author} {\bibinfo {author} {\bibfnamefont {G.}~\bibnamefont
  {Piermarini}}, \bibinfo {author} {\bibfnamefont {S.}~\bibnamefont {Block}},\
  and\ \bibinfo {author} {\bibfnamefont {J.}~\bibnamefont {Barnett}},\
  }\bibfield  {title} {\bibinfo {title} {Hydrostatic limits in liquids and
  solids to 100 kbar},\ }\href@noop {} {\bibfield  {journal} {\bibinfo
  {journal} {J. Appl. Phys.}\ }\textbf {\bibinfo {volume} {44}},\ \bibinfo
  {pages} {5377} (\bibinfo {year} {1973})}\BibitemShut {NoStop}%
\bibitem [{\citenamefont {Olinger}\ and\ \citenamefont
  {Halleck}(1975)}]{olinger1975compression}%
  \BibitemOpen
  \bibfield  {author} {\bibinfo {author} {\bibfnamefont {B.}~\bibnamefont
  {Olinger}}\ and\ \bibinfo {author} {\bibfnamefont {P.~M.}\ \bibnamefont
  {Halleck}},\ }\bibfield  {title} {\bibinfo {title} {Compression and bonding
  of ice vii and an empirical linear expression for the isothermal compression
  of solids},\ }\href@noop {} {\bibfield  {journal} {\bibinfo  {journal} {J.
  Chem. Phys.}\ }\textbf {\bibinfo {volume} {62}},\ \bibinfo {pages} {94}
  (\bibinfo {year} {1975})}\BibitemShut {NoStop}%
\bibitem [{\citenamefont {Xie}\ \emph {et~al.}(2017)\citenamefont {Xie},
  \citenamefont {Zhang},\ and\ \citenamefont {Liu}}]{xie2017graphite}%
  \BibitemOpen
  \bibfield  {author} {\bibinfo {author} {\bibfnamefont {Y.-P.}\ \bibnamefont
  {Xie}}, \bibinfo {author} {\bibfnamefont {X.-J.}\ \bibnamefont {Zhang}},\
  and\ \bibinfo {author} {\bibfnamefont {Z.-P.}\ \bibnamefont {Liu}},\
  }\bibfield  {title} {\bibinfo {title} {Graphite to diamond: origin for
  kinetics selectivity},\ }\href@noop {} {\bibfield  {journal} {\bibinfo
  {journal} {JACS}\ }\textbf {\bibinfo {volume} {139}},\ \bibinfo {pages}
  {2545} (\bibinfo {year} {2017})}\BibitemShut {NoStop}%
\bibitem [{\citenamefont {Stojkovic}\ \emph {et~al.}(2003)\citenamefont
  {Stojkovic}, \citenamefont {Zhang}, \citenamefont {Lammert},\ and\
  \citenamefont {Crespi}}]{stojkovic2003collective}%
  \BibitemOpen
  \bibfield  {author} {\bibinfo {author} {\bibfnamefont {D.}~\bibnamefont
  {Stojkovic}}, \bibinfo {author} {\bibfnamefont {P.}~\bibnamefont {Zhang}},
  \bibinfo {author} {\bibfnamefont {P.~E.}\ \bibnamefont {Lammert}},\ and\
  \bibinfo {author} {\bibfnamefont {V.~H.}\ \bibnamefont {Crespi}},\ }\bibfield
   {title} {\bibinfo {title} {Collective stabilization of hydrogen
  chemisorption on graphenic surfaces},\ }\href@noop {} {\bibfield  {journal}
  {\bibinfo  {journal} {Phys. Rev. B}\ }\textbf {\bibinfo {volume} {68}},\
  \bibinfo {pages} {195406} (\bibinfo {year} {2003})}\BibitemShut {NoStop}%
\end{thebibliography}

\end{document}